 \documentclass[final,5p,times,twocolumn]{elsarticle}

\usepackage[T1]{fontenc}
\usepackage{microtype}
\usepackage{graphicx}
\usepackage{dsfont}
\usepackage{color}
\usepackage{siunitx}
\usepackage{caption}
\usepackage{amssymb}
\usepackage{amsmath}
\usepackage{url} 
\usepackage{hyperref}
\usepackage{orcidlink}

\usepackage[switch]{lineno} 

\journal{Astroparticle Physics}
\newcommand{\Londrina}{Departamento de F\'isica, 
	Universidade Estadual de Londrina, 
	Londrina,
	Brazil}

\newcommand{\PUCR}{Department of Physics, 
	Pontif\'icia Universidade Cat\'olica do Rio de Janeiro, 
	Rio de Janeiro, Brazil}

\newcommand{\TechnicoIBB}{iBB, 
	Instituto Superior Tecnico, 
	Universidade de Lisboa,
	Lisbon, 
	Portugal}
    
\newcommand{\TechnicoIDMEC}{IDMEC, 
	Instituto Superior Tecnico, 
	Universidade de Lisboa,
	Lisbon, 
	Portugal}

\newcommand{\TechnicoCtwoTN}{CT2N, 
	Instituto Superior Tecnico, 
	Universidade de Lisboa,
	Lisbon, 
	Portugal}
    
\newcommand{\Queens}{Department of Physics, 
	Engineering Physics \& Astronomy, 
	Queen's University, Kingston, 
	Canada}

\newcommand{\Prague}{Institute of Particle and Nuclear Physics,
	Charles University,
	Prague, 
	Czech Republic}

\newcommand{\IJCLabSaclay}{Universit\'e Paris-Saclay, 
	CNRS/IN2P3, 
	IJCLab, 
	Orsay, 
	France}

\newcommand{\LPtwoI}{Universit\'e de Bordeaux, 
	CNRS, 
	LP2I Bordeaux, 
	Gradignan, 
	France}

\newcommand{\SUBA}{Nantes Universit\'e, 
	IMT-Atlantique, 
	CNRS, 
	Subatech,
	Nantes, 
	France}
	
\newcommand{\CPPM}{Universit\'e de Aix Marseille, 
	CNRS, 
	CPPM, 
	Marseille, 
	France}

\newcommand{\LNCA}{LNCA Underground Laboratory, 
	CNRS, 
	EDF Chooz Nuclear Reactor, 
	Chooz, France}

\newcommand{\FerraraUni}{Dipartimento di Fisica e Scienze della Terra, 
	Universit\`{a} di Ferrara, 
	Ferrara, 
	Italy}
\newcommand{\FerraraINFN}{INFN, 
	Sezione di Ferrara, 
	Ferrara, 
	Italy}

\newcommand{\Padovauni}{Dipartimento di Fisica e Astronomia, 
	Universit\`{a} di Padova, 
	Padova, Italy}
\newcommand{\Padova}{INFN, 
	Sezione di Padova, 
	Padova, Italy}

\newcommand{\MainzA}{Johannes Gutenberg-Universit\"{a}t Mainz,
	Institut f\"{u}r Physik, 
	Mainz, Germany}

\newcommand{\CIEMAT}{CIEMAT, 
	Centro de Investigaciones Energ\'{e}ticas, Medioambientales y Tecnol\'{o}gicas, 
	Madrid, Spain}

\newcommand{\UniZar}{Centro de Astropart\'{\i}culas y F\'{\i}sica de Altas Energ\'{\i}as (CAPA),
 	Universidad de Zaragoza, 
	Zaragoza, Spain}

\newcommand{\DIPC}{Donostia International Physics Center, 
	Basque Excellence Research Centre, 
	San Sebasti\'an/Donostia, 
	Spain}

\newcommand{\RCNS}{RCNS, 
	Tohoku University, 
	Sendai, Japan}

\newcommand{\Sussex}{Department of Physics and Astronomy, 
	University of Sussex, 
	Brighton, 
	United Kingdom}

\newcommand{\ImpCol}{Department of Chemistry, 
	Imperial College London, 
	London, 
	United Kingdom}
	
\newcommand{\RAL}{Rutherford Appleton Laboratory, 
	Didcot,
	Oxford,
	United Kingdom}

\newcommand{\UCI}{Department of Physics and Astronomy, 
	University of California at Irvine, 
	Irvine, 
	CA, 
	USA}

\newcommand{\UniPennPhys}{Department of Astronomy and Astrophysics, 
	Pennsylvania State University, 
	University Park, 
	PA,
	USA}
\newcommand{\UniPennAstro}{Department of Physics, 
	Pennsylvania State University, 
	University Park, 
	PA,
	USA}

\newcommand{\UniMichigan}{Department of Nuclear Engineering and Radiological Sciences,
 	University of Michigan, 
	Ann Arbor, 
	MI,
	USA}
					
\newcommand{\BNL}{Brookhaven National Laboratory, 
	Upton, 
	NY,
	USA}
\newcommand{\red}{\textcolor{red}}

\begin{document}

\begin{frontmatter}



\title{COCOA: a compact Compton camera for astrophysical observation of MeV-scale gamma rays\\\vspace{1.2em}\large{LiquidO collaboration}\vspace{-0.8em}}

% AUTHORS
\author[u]{S.\,R.\,Soleti\orcidlink{0000-0002-5526-1414}\corref{cor1}}\ead{roberto.soleti@dipc.org}
\cortext[cor1]{Corresponding author}
\author[u]{J.\,J.\,G\'omez-Cadenas\orcidlink{0000-0002-8224-7714}}

\author[z]{J.\,Apilluelo}
\author[b]{L.\,Asquith}
\author[b]{E.\,F.\,Bannister}
\author[k1]{N.\,P.\,Barradas}
\author[b]{C.\,L.\,Baylis}
\author[p]{J.\,L.\,Beney}
\author[k2]{M.\,Berberan\,e\,Santos}
\author[p]{X.\,de\,la\,Bernardie}
\author[b]{T.\,J.\,C.\,Bezerra\orcidlink{0000-0002-0424-7903}} 
\author[p]{M.\,Bongrand} 
\author[q]{C.\,Bourgeois}
\author[q]{D.\,Breton}
\author[n]{J.\,Busto}
\author[q]{K.\,Burns}
\author[q,c]{A.\,Cabrera\orcidlink{0000-0001-5713-3347}}
\author[p]{A.\,Cadiou}
\author[l]{E.\,Calvo}
\author[b]{M.\,de\,Carlos\,Generowicz}
\author[f]{E.\,Chauveau}
\author[b]{B.\,J.\,Cattermole}
\author[h]{M.\,Chen}
\author[i]{P.\,Chimenti} 
\author[x1,x2]{D.\,F.\,Cowen}
\author[b]{S.\,Kr.\,Das}
\author[r1]{S.\,Dusini\orcidlink{0000-0002-1128-0664}} 
\author[b]{A.\,Earle}
\author[k1]{M.\,Felizardo}
\author[i]{C.\,Frigerio\,Martins}
\author[z]{J.\,Gal\'an}
\author[z]{J.\,A.\,Garc\'ia}
\author[q]{R.\,Gazzini}
\author[b]{A.\,Gibson-Foster}
\author[m1]{C.\,Girard-Carillo}
\author[b]{W.\,C.\,Griffith}
\author[p]{M.\,Guiti\`ere}
\author[p]{F.\,Haddad}
\author[b]{J.\,Hartnell} 
\author[d]{A.\,Holin}
\author[z]{I.\,G.\,Irastorza}
\author[a]{I.\,Jovanovic\orcidlink{0000-0003-0573-3150}}
\author[k1]{A.\,Kling}
\author[m1]{L.\,Koch\orcidlink{0000-0002-2966-7461}}
\author[b]{P.\,Lasorak}
\author[q,c]{J.\,F.\,Le\,Du}
\author[p]{F.\,Lefevre}
\author[q]{P.\,Loaiza}
\author[b]{J.\,A.\,Lock}
\author[z]{G.\,Luz\'on}
\author[q]{J.\,Maalmi}
\author[j]{J.\,P.\,Malhado}
\author[e1,e2]{F.\,Mantovani}
\author[k1]{J.\,G.\,Marques}
\author[f]{C.\,Marquet} 
\author[z]{M.\,Mart\'inez}
\author[l]{D.\,Navas-Nicol\'as\orcidlink{0000-0002-2245-4404}}
\author[t]{H.\,Nunokawa} 
\author[g]{J.\,P.\,Ochoa-Ricoux\orcidlink{0000-0001-7376-5555}} 
\author[k2]{T.\,Palmeira}
\author[l]{C.\,Palomares} 
\author[d]{D.\,Petyt}
\author[p]{P.\,Pillot}
\author[f]{A.\,Pin}
\author[b]{J.\,C.\,C.\,Porter} 
\author[f]{M.\,S.\,Pravikoff\orcidlink{0000-0002-7088-4126}}
\author[d]{S.\,Richards}
\author[k2]{N.\,Rodrigues}
\author[f]{M.\,Roche}
\author[y]{R.\,Rosero}
\author[s]{B.\,Roskovec}
\author[q]{N.\,Roy}
\author[z]{M.\,L.\,Sarsa}
\author[r1,r2]{A.\,Serafini}
\author[d]{C.\,Shepherd-Themistocleous}
\author[b]{W.\,Shorrock\orcidlink{0000-0002-7221-1910}}
\author[k3]{M.\,Silva}
\author[q]{L.\,Simard}
\author[p]{D.\,Stocco}
\author[e1,e2]{V.\,Strati}
\author[p]{J.\,S.\,Stutzmann}
\author[v]{F.\,Suekane}
\author[b]{N.\,Tuccori\orcidlink{0000-0002-2868-5887}}
\author[l]{A.\,Verdugo}
\author[p]{B.\,Viaud}
\author[m1]{S.\,M.\,Wakely\orcidlink{0000-0002-2919-8159}}
\author[x2]{G.\,Wendel}
\author[a]{A.\,S.\,Wilhelm\orcidlink{0000-0002-0664-0477}}
\author[b]{A.\,W.\,R.\,Wong}
\author[y]{M.\,Yeh}
\author[p]{F.\,Yermia}

%
%AUTHORS: 98
%
% For LiquidO-I 
%
%
% INSTITUTIONS: -- ordered alphabetically
%
%CURRENT:
%
\affiliation[a]{organization=\UniMichigan} %UPDATE: JUNE 2023 - yes 14
%B
\affiliation[b]{organization=\Sussex} %UPDATE: JUNE 2023? 22 !!!!!!
%C
\affiliation[c]{organization=\LNCA} %UPDATE: JUNE 2023 - yes 10
%D:
\affiliation[d]{organization=\RAL} 
%E
%F:
\affiliation[e1]{organization=\FerraraINFN} %UPDATE: JUNE 2023 - yes 5
\affiliation[e2]{organization=\FerraraUni} %UPDATE: JUNE 2023 - yes 5
%G
\affiliation[f]{organization=\LPtwoI} %UPDATE: JUNE 2023 - yes 12
%H
%I
\affiliation[g]{organization=\UCI} %UPDATE: JUNE 2023 - yes 8
%J
%K
\affiliation[h]{organization=\Queens} %UPDATE: JUNE 2023 - yes 19
%L
\affiliation[i]{organization=\Londrina} %UPDATE: JUNE 2023 - yes 11
\affiliation[j]{organization=\ImpCol} %UPDATE: JUNE 2023 - yes 7 !!!!!!
\affiliation[k1]{organization=\TechnicoCtwoTN} %UPDATE: JUNE 2023 - yes 23 !!!!!!
\affiliation[k2]{organization=\TechnicoIBB} %UPDATE: JAN 2025 - yes 23 !!!!!!
\affiliation[k3]{organization=\TechnicoIDMEC} %UPDATE: JAN 2025 - yes 23 !!!!!!
%M:
\affiliation[l]{organization=\CIEMAT} %UPDATE: JUNE 2023 - yes 3
\affiliation[m1]{organization=\MainzA} %UPDATE: JUNE 2023 - yes 13
% \affiliation[m2]{organization=\MainzB} %UPDATE: JUNE 2023 - yes 13
\affiliation[n]{organization=\CPPM} 
%N:
\affiliation[p]{organization=\SUBA} %UPDATE: JUNE 2023? 21 !!!!!!
%O
\affiliation[q]{organization=\IJCLabSaclay} %UPDATE: JUNE 2023 - yes 6
%P
\affiliation[r1]{organization=\Padova} %UPDATE: JUNE 2023 - yes 16
\affiliation[r2]{organization=\Padovauni} %UPDATE: JUNE 2023 - yes 16
\affiliation[s]{organization=\Prague} %UPDATE: JUNE 2023 - yes 2
%Q
%R
\affiliation[t]{organization=\PUCR} %UPDATE: JUNE 2023 - yes 18
%S
\affiliation[u]{organization=\DIPC} %UPDATE: JUNE 2023 - yes 4 !!!!!!
\affiliation[v]{organization=\RCNS} %UPDATE: JUNE 2023 - yes 20
%T
%U
\affiliation[x1]{organization=\UniPennPhys} %UPDATE: JUNE 2023 - yes 17
\affiliation[x2]{organization=\UniPennAstro} %UPDATE: JUNE 2023 - yes 17
\affiliation[y]{organization=\BNL} %UPDATE: JUNE 2023 - yes 1 !!!!!!
%V
%W
%Z
\affiliation[z]{organization=\UniZar} %UPDATE: JUNE 2023 - yes 24
%
% APPENDED -- not members HISTORY
%
% \affiliation[1]{organization=\MPIK} %UPDATE: JUNE 2023 - yes 25 !!!!!!
% \affiliation[2]{organization=\APC} %UPDATE: JUNE 2023 - yes 26 !!!!!!

\begin{abstract}

COCOA (COmpact COmpton cAmera) is a next-generation gamma-ray telescope designed for astrophysical observations in the MeV energy range. The detector comprises a scatterer volume employing the LiquidO detection technology and an array of scintillating crystals acting as absorber. Surrounding plastic scintillator panels serve as a veto system for charged particles. The detector’s compact, scalable design enables flexible deployment on microsatellites or high-altitude balloons. Gamma rays at MeV energies have not been well explored historically (the so-called ``MeV gap") and COCOA has the potential to improve the sensitivity in this energy band.
\end{abstract}




\end{frontmatter}

% \tableofcontents

% \linenumbers


\section{Introduction}
MeV gamma-ray observations are crucial for addressing many unresolved questions in astrophysics. Notable examples include the definitive identification of the origin of cosmic rays through the detection of nuclear de-excitation line emissions in the few MeV range~\cite{benhabiles2013excitation} and the distribution of Al-26 nuclei, observed at 1.8 MeV, which reveals the sites of nucleosynthesis within our Galaxy~\cite{knodlseder1999implications}. Additionally, MeV gamma rays may be produced alongside gravitational waves from neutron star mergers, making their detection an important component of multi-messenger astrophysics~\cite{LIGOScientific:2017ync}.

\textcolor{black}{The observation of astrophysical gamma rays in this energy range has been pioneered by the Compton telescope COMPTEL~\cite{schonfelder1993instrument}. Most notably, this experiment detected an inner Galactic emission in the 1–30 MeV energy range, the origin of which has remained unresolved since its discovery~\cite{tsuji2023mev}. If the galactic diffuse emission model of
Fermi-LAT is extrapolated to the MeV energy range, there is an apparent excess component to account for the COMPTEL emission, dubbed the \emph{COMPTEL excess}. A possible explanation for this excess might be the presence of annihilation or decay of dark matter~\cite{Boddy:2015efa}. Despite this, the sensitivity of past and current Compton telescopes in this energy band is still low, compared to other energy regions~\cite{kierans2024compton}.}

\begin{figure}[htbp]
\centering
\includegraphics[width=0.99\linewidth]{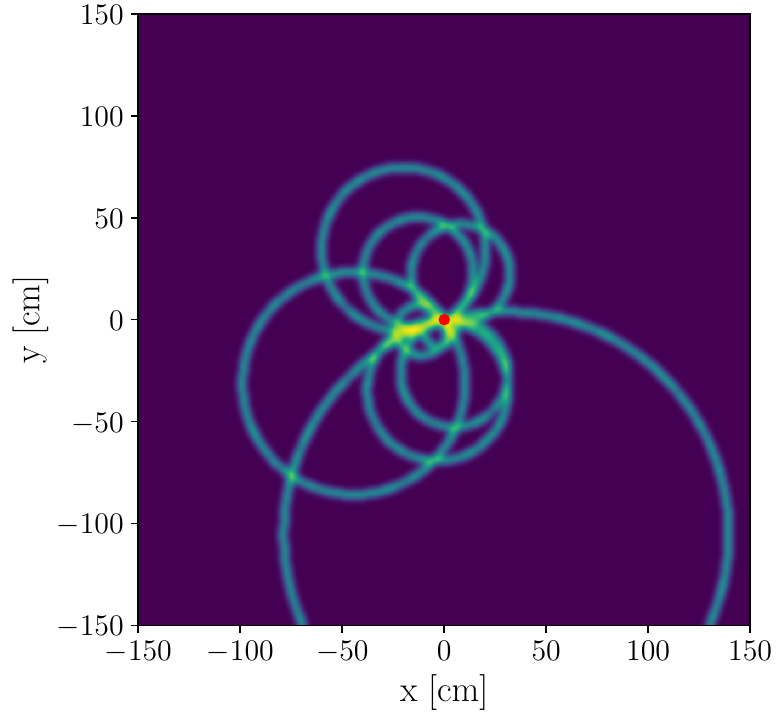}

\caption{Back-projected event circles from ten gamma rays of 1~MeV generated above the COCOA detector at $(0,0)$, impinging at different angles. The red dot corresponds to the true starting position of the gamma. The blurring is caused by the finite angular resolution of the instrument (see section~\ref{sec:arm}).}\label{fig:backprojected}
\end{figure}

Traditionally, a Compton telescope is composed of one or more scatterer layers, made of low-Z material to maximize the probability of Compton scattering, and an absorber volume, where gamma rays are photo-absorbed and their final energy is measured. Thus, for a two-site event, the scatter angle $\theta$ can then be derived from the Compton equation as:

\begin{equation}\label{eq:compton}
    \cos\theta = 1 - \frac{m}{E_2} + \frac{m}{E_1+E_2},
\end{equation}
where $m$ is the mass of the electron, $E_1$ is the energy deposited in the scatterer and $E_2$ is the energy deposited in the absorber. The angle $\theta$ identifies an event circle in the sky: when multiple gammas from the same source are detected, the overlap of each event circle allows to locate the position of the source, as shown in fig.~\ref{fig:backprojected}. 

\textcolor{black}{The efficiency of this kind of detector is typically low, less than 1\% for COMPTEL~\cite{kierans2024compton}, primarily due to its large physical separation between the scatterer and the absorber. This configuration results in a small solid angle for the absorber as seen from the scatterer. To overcome this limitation, most modern \textcolor{black}{concepts} adopt more compact geometries using three-dimensional position-sensitive detectors, such as cryogenic germanium detectors~\cite{Tomsick:2021H5}, silicon strip detectors~\cite{Bloser:2001dj, e-ASTROGAM:2016bph}, or time projection chambers~\cite{Aramaki:2019bpi, Shutt:2025xvc}. These designs enhance the probability of detecting one or more scatters within an active volume. In the case of multiple scatters, eq.~\eqref{eq:compton} can then be scaled to $n>2$ interactions:}

\begin{equation}
    \cos\theta = 1 - \frac{m}{\sum_{i=2}^n E_i} + \frac{m}{\sum_{i=1}^n E_i}.
\end{equation}

\textcolor{black}{However, a compact layout introduces new challenges: it makes it challenging to measure time-of-flight (TOF) information -- an important tool for background suppression in COMPTEL~\cite{Weidenspointner:2000aq} -- and place stringent requirements on spatial resolution in order to maintain good angular resolution. Therefore, beyond increasing interaction probability, modern designs must also ensure sufficient angular resolution and effective background rejection.}

Above $\mathcal{O}(10~\mathrm{MeV})$, pair-production becomes the dominant process contributing to the gamma cross-section. Three-dimensional detectors can typically reconstruct the energy and the trajectory of the $e^+e^-$ pairs. This information is then used to determine the position of the gamma source in the sky.

The excellent performances and capabilities of these technologies, however, come at the expense of significant complications (e.g., cryogenics, high segmentation, large number of channels) and cost. 

The key feature of the COCOA detector is the adoption of the LiquidO technology~\cite{LiquidO:2019mxd} for the scatterer, which allows three-dimensional positioning of the interactions inside a volume filled with an \emph{opaque medium}~\cite{Buck:2019tsa, Tang:2020xoy}. Typically, this medium can be an opaque liquid scintillator (OLS), for which multiple formulations are under development, each tailored to different operational environments~\cite{cabrera_2024_10647143}. While formulations based on wax or other temperature-sensitive materials have been successfully demonstrated on Earth, their suitability for space conditions remains to be validated. Nevertheless, current knowledge and ongoing R\&D efforts~\cite{polupan2023peculiarities, Schoppmann:2022hst, Wagner:2018ajx} indicate no fundamental showstoppers for the development of space-qualified OLS formulations. 

Due to the short scattering length and long attenuation length of such materials, scintillation and Cherenkov photons travel only a short distance between scatters, resulting in \emph{stochastic confinement}~\cite{LiquidO:2025qia}.
These photons can then be captured by a grid of wavelength-shifting (WLS) optical fibers threading the detector volume, which subsequently re-emit them at longer wavelengths. A portion of the re-emitted photons propagates through the fibers via total internal reflection and is finally detected by photosensors placed at one or both ends of the fibers. This design achieves virtual voxelization without requiring a physical segmentation of the detector volume, as in the case of, e.g., the SoLid~\cite{SoLid:2020cen} and SuperFGD detectors~\cite{Blondel:2017orl}.

Although the technology was originally developed for antineutrino detection~\cite{anatael_cabrera_2022_7504162, cabrera_2023_10049846}, its potential applications are also being explored in other fields, including positron emission tomography (PET) scanners~\cite{anatael_cabrera_2022_7556760}. Interestingly, this application shares several key requirements with gamma-ray telescopes — specifically, the ability to reconstruct Compton interactions, along with \textcolor{black}{percent-level} energy resolution, \textcolor{black}{millimeter-scale} spatial resolution, and high detection efficiency at the MeV scale.

In particular, LiquidO enables sub-centimeter spatial resolution~\cite{LiquidO:2024piw} and can achieve an energy resolution of approximately $5\%/\sqrt{\mathrm{MeV}}$~\cite{LiquidO:2019mxd}, thanks to the high light yield of the scintillator (typically $\sim10^5$~photons/MeV) and the good light collection efficiency of the technology (in the order of 10\%).
This choice significantly increases the effective area of the telescope and enables the detection of events with multiple interactions inside the scatterer.

Notably, the number of channels scales with the surface area rather than with the volume, in contrast with Compton telescopes employing, e.g., semiconductor detectors. COCOA baseline design requires approximately three thousand electronics channels, which is two orders of magnitude less than e-ASTROGAM~\cite{e-ASTROGAM:2016bph}, based on silicon strips, and 3-5 times less than GRAMS~\cite{Aramaki:2019bpi}, a liquid argon time projection chamber. 

Most interestingly, recent advancements in reusable rocket technology allow sending small and medium-scale satellites to low-earth orbit (LEO) with a \textcolor{black}{launch} cost that is an order of magnitude lower than one or two decades ago~\cite{jones2018recent}. The compact dimensions of COCOA and its relatively low weight make it an attractive candidate for a microsatellite mission.%, with the potential to achieve sensitivities comparable to next-generation experiments~\cite{e-ASTROGAM:2016bph, tomsick2019compton} at a fraction of the cost.

This document is organized as follows. Section~\ref{sec:apparatus} describes the experimental apparatus of COCOA, while the expected performances and sensitivities are detailed in section~\ref{sec:performances}. Two possible mission profiles are described in section~\ref{sec:mission}. Finally, section~\ref{sec:conclusions} contains a summary of the results and future prospects.

\section{Experimental apparatus}\label{sec:apparatus}

\begin{figure}[htbp]
\centering
\includegraphics[width=\linewidth]{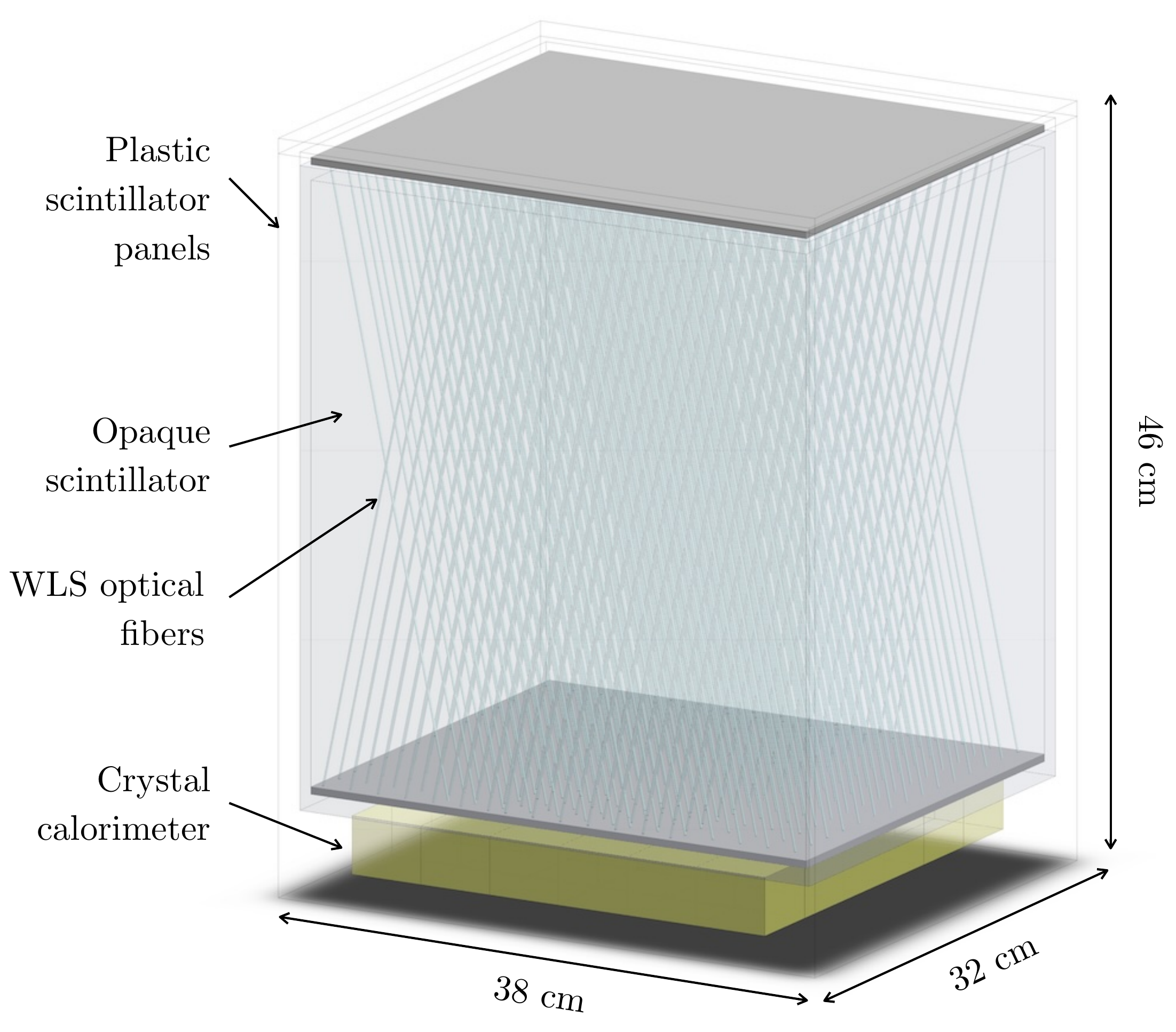}

\caption{CAD drawing of the COCOA detector. It comprises an opaque scintillator intersected by a grid of wavelength-shifting fibers, with a crystal calorimeter placed on the bottom. It is surrounded by plastic scintillator panels to reject the charged particles background.}\label{fig:cocoa}
\end{figure}

The COCOA detector, shown in fig.~\ref{fig:cocoa}, is divided into two main parts, a \emph{scatterer} and an \emph{absorber}. These two elements are surrounded by plastic scintillator panels to veto the charged particles background. It has a total size of $38\times32\times46$~cm$^3$ and a total weight of approximately 50~kg, excluding power supply and data acquisition systems. The baseline system specifications are detailed in the following sections and summarized in table~\ref{tab:specifications}.

\begin{figure*}[ht!]
\centering
\includegraphics[width=1\linewidth]{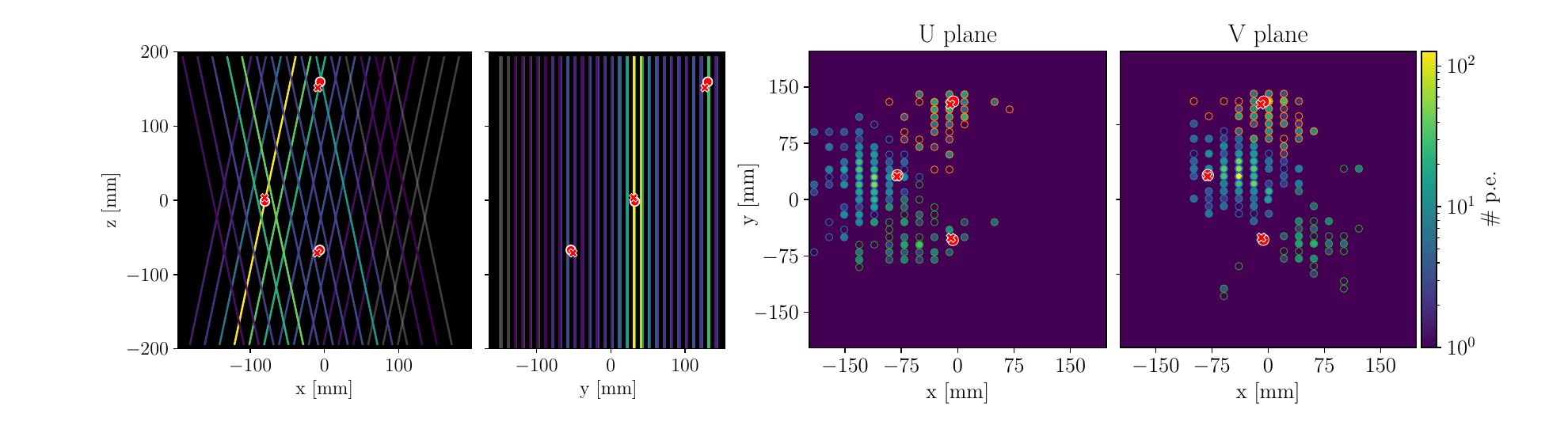}
\caption{\textcolor{black}{Simulated event display of a gamma particle with an energy of 2.5 MeV scattering three times inside the opaque scintillator. The red dots and red crosses correspond to the true and reconstructed scattering positions, respectively. On the U and V projections, the inner color of each fiber represents the number of photoelectrons detected, shown on a logarithmic scale, while the edge color corresponds to the three reconstructed clusters. The fibers are inclined at $12^{\circ}$, so the centroids of the clusters in the U and V planes (not shown in the figure) do not directly correspond to the true interaction positions. Instead, the transverse coordinates are estimated using a charge-weighted average of the centroids in the two planes. The longitudinal coordinate $z$ can then be extracted from the distance between the two centroids in the transverse plane, knowing the inclination of the fibers.}}\label{fig:event_scatterer}
\end{figure*}

\begin{table}[ht!]
\centering
\begin{tabular}{l|c}
\hline
Parameter&Specification\\
\hline
Energy band & 0.2 -- 100 MeV \\
Weight & 50~kg\\
Dimensions & $38\times32\times46$~cm$^3$\\
Angular resolution & $4.0^{\circ}$ at 1 MeV \\ 
Electronics channels & 3376 \\
Scatterer material & Opaque scintillator \\
Absorber material & LaBr$_3$(Ce) or CsI(Tl)\\
$\Delta \vec{x}/\vec{x}$ scatterer & $(8, 6, 15)$~mm at 1~MeV \\
$\Delta \vec{x}/\vec{x}$ absorber & $(2, 2, 2)$~mm at 1~MeV \\
$\Delta E /E$ scatterer & $5\%/\sqrt{E / \mathrm{MeV}}$\\
$\Delta E /E$ absorber & $2.5\%/\sqrt{E / \mathrm{MeV}}$\\
Satellite power & $210$~W\\
Satellite telemetry & 20~GB/day\\
Satellite attitude & 30"\\
Mission duration & 1+ years (satellite)\\
\hline
\end{tabular}
\caption{Specifications of the COCOA systems and mission. Angular, energy and spatial resolutions are intended as FWHM.\label{tab:specifications}}
\end{table}

\subsection{Scatterer}

The scatterer volume, of dimensions $36\times30\times38$~cm$^3$, is filled with 41~L of opaque liquid scintillator. In the baseline design of COCOA we adopt the NoWaSH cocktail. This consists of a mix of a solvent, a fluor and a wax, to ensure the opacity of the medium. A typical combination is represented by linear alkylbenzene (LAB) doped with 0.3\% diphenyloxazole (PPO) and mixed with paraffin~\cite{Buck:2019tsa}. The amount of wax can be adjusted to achieve the desired scattering length and light yield, with typical values ranging from 2\% to 20\% by weight. This mix has already been tested by the LiquidO collaboration~\cite{Bezerra:2024frp,LiquidO:2025qia} and its production in the quantities needed by COCOA is not a challenge. A possible alternative is represented by water-based opaque liquid scintillators (WbLS), where the opacity is achieved by adding water to the mix and a better Cherenkov/scintillation separation can be achieved~\cite{LiquidO:2024piw}. 

The light produced by the scintillator is collected by round WLS fibers with a diameter of 1~mm, threading through the medium along the $z$-axis. One end of each fiber is coupled to a silicon photomultiplier (SiPM) with an active area of $1.3\times1.3$~mm$^2$, while the other end is coated with an aluminum mirror, which offer approximately 75\% reflectivity~\cite{Saraiva:2004cn}. 

The fibers are arranged into two $30\times15$ matrices, one at $+12^{\circ}$ with respect to the $z-y$ plane, called the V plane, and one at $-12^{\circ}$, called the U plane. The pitch in the $x$ and $y$ directions is 2~cm and 1~cm, respectively. Along the $y$-axis, the U plane is shifted by 1~mm to avoid collisions between fibers. This layout corresponds to a manageable number of channels (900), with modest power and data acquisition needs. The position of the interaction can be extracted from the position of the photosensors collecting the light, as exemplified in figure~\ref{fig:event_scatterer} and detailed in section~\ref{sec:performances}.

\begin{figure}[ht!]
\centering
\includegraphics[width=0.99\linewidth]{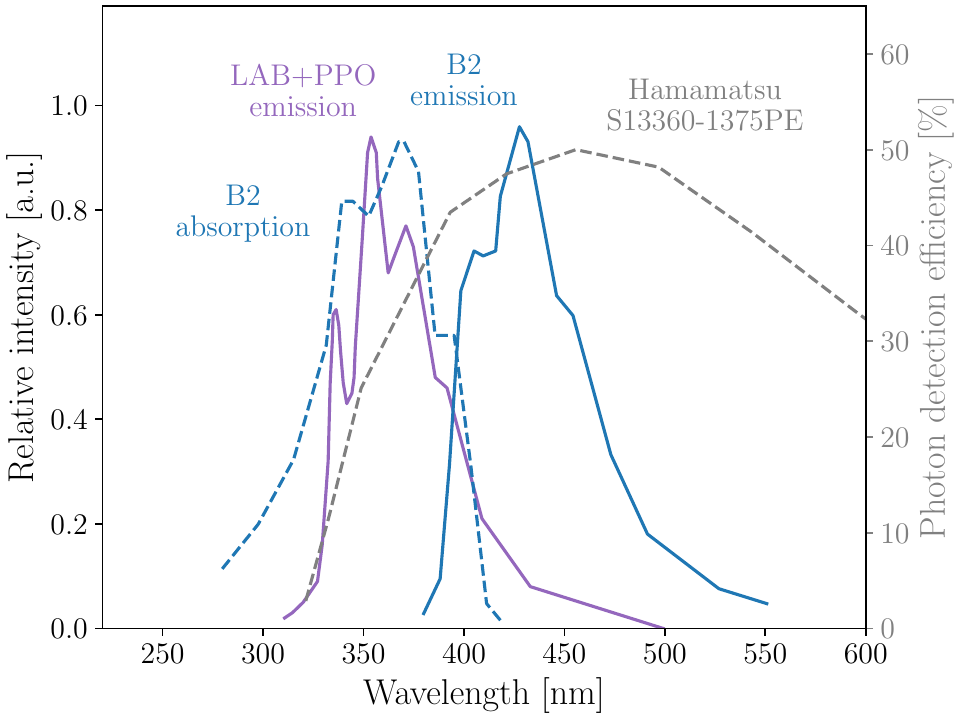}
\caption{Emission and absorption spectra for the B2 fibers (in blue), compared with the LAB+PPO emission spectrum (in purple) and the Hamamatsu MPPC S13360-1375PE PDE (in gray, right axis).}\label{fig:mppc}
\end{figure}

The chosen candidate for the WLS fiber is the Kuraray B2 model~\cite{kuraray}, whose absorption spectrum closely matches the LAB+PPO emission spectrum~\cite{Kaptanoglu:2018sus} and has, in turn, an emission spectrum compatible with the photon detection efficiency (PDE) of the Hamamatsu MPPC S13360-1375PE~\cite{Hamamatsu}, as shown in fig.~\ref{fig:mppc}. 

The SiPMs can be amplified, shaped and digitized with the BETA application-specific integrated circuit (ASIC)~\cite{sanmukh2024low}. This chip was initially developed for the High Energy Cosmic-Radiation Detection (HERD) facility onboard the Future Chinese Space Station ~\cite{gargano2021high}. BETA has a power consumption of approximately 1~mW/channel and a maximum rate of 10~kHz. Measurements with $3\times3$~mm$^2$ Hamamatsu MPPCs showed a time resolution of 400 ps FWHM for signals of 10 photoelectrons.

\subsection{Absorber}

The absorber, placed below the scatterer volume at a distance of 5~mm from its bottom face, consists of a crystal calorimeter with an area of $30\times30$~cm${^2}$ and a thickness of 2 radiation lengths, which depends on the material. Here, a high-density, high-Z scintillator with a good light yield is required. The ideal choice is represented by LaBr$_3$(Ce), a modern inorganic crystal with excellent energy resolution (better than 4\% at 662~keV~\cite{Giaz:2013hna}). A cost-effective alternative can be represented by thallium-doped cesium iodide, CsI(Tl), which provides an energy resolution below 5\% at 662~keV~\cite{e-ASTROGAM:2016bph} and has a solid track-record in astroparticle physics~\cite{Fermi-LAT:2009ihh}. The crystals will be wrapped on five sides with ESR Vikuiti film~\cite{3M}, which provides reflectivity above 98\% in the visible spectrum. This is a non-metallic polymer that has already been successfully used in space in the Fermi/LAT calorimeter~\cite{atwood2009large}.

\begin{figure}[ht!]
\centering
\includegraphics[width=0.99\linewidth]{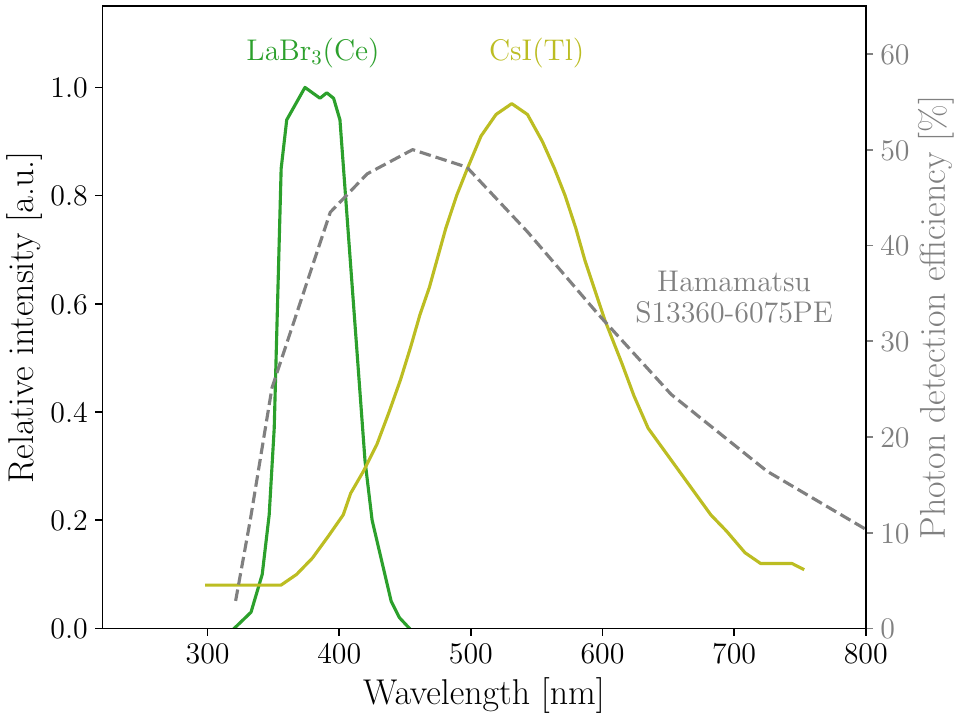}
\caption{Emission spectrum of the scintillating crystals LaBr$_3$(Ce) (in green) and CsI(Tl) (in yellow) compared with Hamamatsu MPPC S13360-6075PE PDE (in gray, right axis). Data taken from vendors ~\cite{Hamamatsu, Advatech_labr3, Advatech_csitl}.}\label{fig:mppc_abs}
\end{figure}

A typical solution for the absorber of a Compton telescopes is represented by a matrix of crystals with a relatively small cross section (a \emph{pixelated} layout), see, e.g., ref.~\cite{Bloser:2001dj}. In this design, the spatial resolution is determined by the pixel size. Therefore, the resolution on the \emph{depth of interaction} (d.o.i.) is typically limited, since the crystals must have at least 2-3 radiation lengths in the thickness dimension (so several centimeters). It is possible to improve this value by further segmenting the detector along the longitudinal direction and place photosensors at different depths. However, this solution introduces significant mechanical complications and an increase in the number of electronics channels.

In COCOA, instead, the calorimeter consists of a matrix of $6\times6$ \emph{monolithic} crystals, each one with an area of $50\times50$~mm$^2$ and read out on one side by a $8\times8$ array of Hamamatsu S13360-6075PE MPPCs, which have a sensitive area of $6\times6$~mm$^2$. The PDE of these photosensors is well matched both to the emission spectrum of LaBr$_3$(Ce)~\cite{Advatech_labr3} and to the one of CsI(Tl)~\cite{Advatech_csitl} (see fig.~\ref{fig:mppc_abs}).

\begin{figure}[ht!]
\centering  
\includegraphics[width=0.99\linewidth]{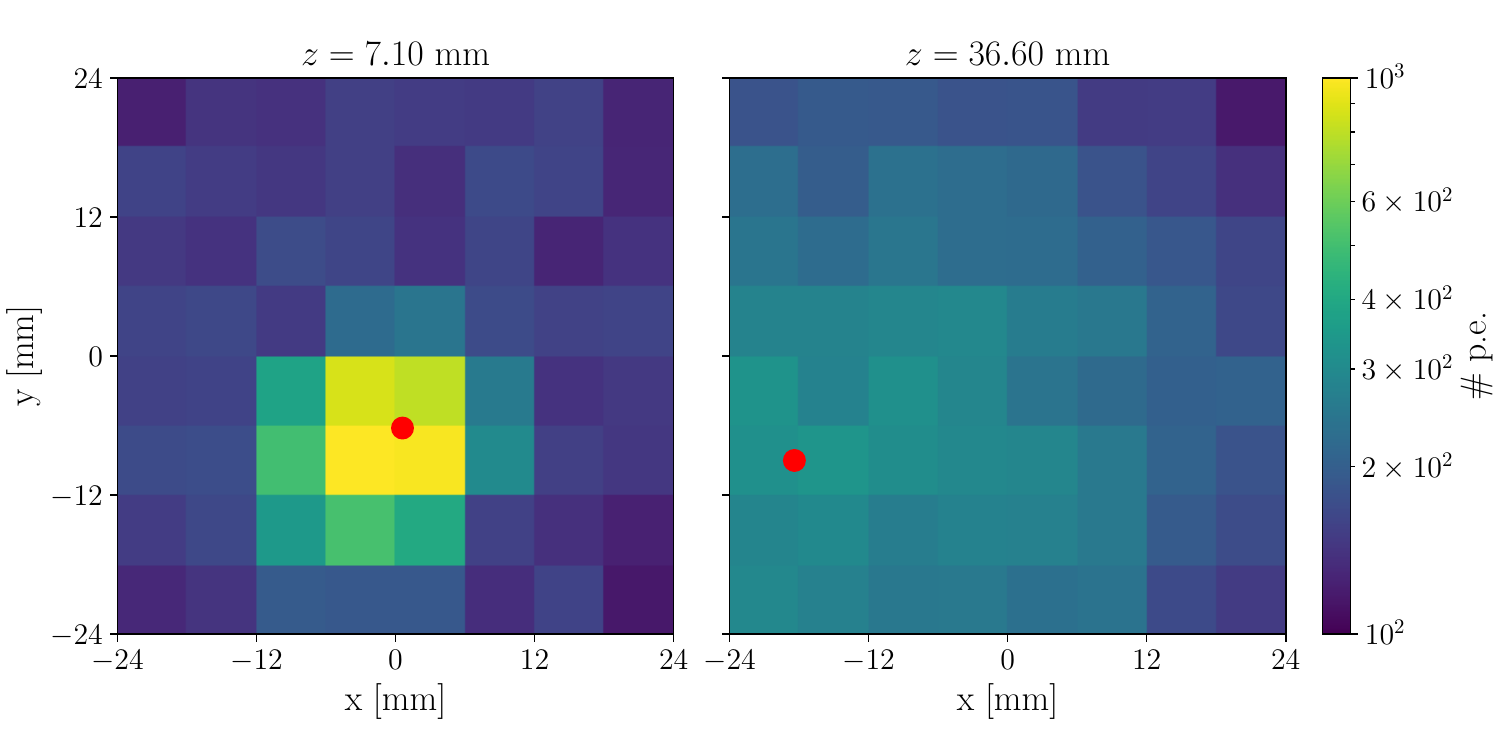}
\caption{Charge collected by the SiPM matrix coupled to a monolithic crystal for two different interactions, one close to the SiPMs plane ($z=7.10$~mm, left) and one near the crystal entrance face ($z=36.60$~mm, right), with the SiPMs plane placed at $z=0$~mm. The red dot corresponds to the simulated interaction point and the color to the number of photoelectrons, in logarithmic scale. The difference between the two patterns is exploited by a CNN to reconstruct the d.o.i. (here the $z$ coordinate).}\label{fig:absorber_evd}
\end{figure}

The advantage of this design choice is that information on the d.o.i. can be extracted from the light spatial distribution: an interaction closer to the sensor will have a narrow distribution, while an interaction closer to the entrance face of the crystal will correspond to a more uniform distribution among the 64 SiPMs, as exemplified in fig~\ref{fig:absorber_evd}. This method, which is actively being explored both for PET scanners~\cite{gonzalez2021evolution} and gamma-ray detectors~\cite{Gostojic:2016joh}, allows to achieve millimeter-scale resolution in all three dimensions within the crystal, as detailed in section~\ref{sec:spatial_res}. A possible drawback of this design choice is the increased pile-up, especially for relatively slow crystals such as CsI(Tl). Thus, its viability needs to be confirmed with a detailed simulation of the expected interaction rate.

The BETA ASIC is the baseline design choice as electronics front-end also for this sub-detector, which will need 2304 electronics channels (one per SiPM). A possible alternative is represented by the VATA64 ASIC, which is adopted by the e-ASTROGAM mission~\cite{e-ASTROGAM:2016bph} and is already space qualified~\cite{bagliesi2011custom}. 

\subsection{Charged particles veto}
The scatterer and absorber volumes are enclosed by segmented plastic scintillator tiles. The four lateral sides of the detector will be covered with 70 tiles of dimensions $460\times20\times5$~mm$^3$, while the top side will have 16 tiles of dimensions $380\times20\times5$~mm$^3$. The bottom side remains open to allow for cabling and detector access. The light produced by the plastic is collected by two WLS fibers placed at the sides of the tile, each one read out by a SiPM, for a total of 172 channels. If necessary, an additional layer of tiles can be installed in a perpendicular orientation to provide bi-dimensional information on the hit position of the charged particle.

This sub-detector, placed in anticoincidence with the scatterer and the absorber, can achieve a background rejection better than $10^{-4}$~\cite{e-ASTROGAM:2016bph}. It has been successfully used in space by the Fermi/LAT~\cite{atwood2009large} and AGILE~\cite{tavani2008agile} missions and is used extensively in particle physics experiments~\cite{Auger:2016tjc, acciarri2021cosmic}. Possible candidates for the plastic scintillator are BC408 or polystyrene-based mixtures containing Diphenylbenzene (PTP) and Bis(5-phenyl-2-oxazolyl)benzene (POPOP).

\begin{figure*}[ht!]
\centering
\includegraphics[width=0.98\linewidth]{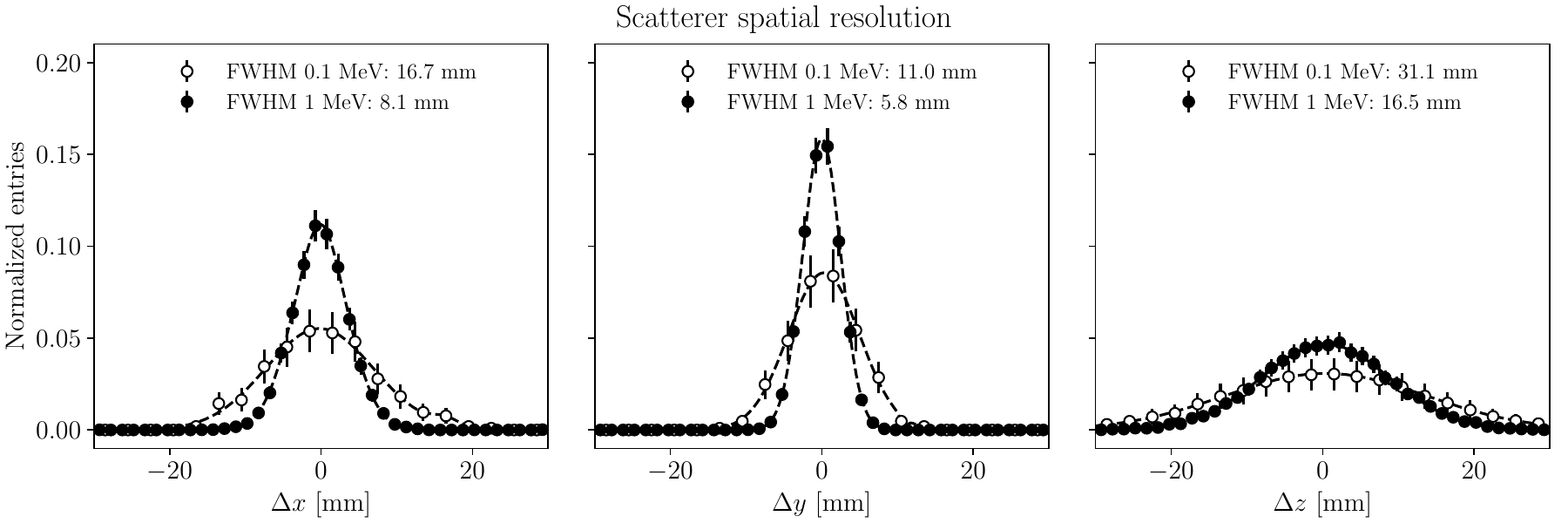}
\caption{Spatial resolution in the three dimensions for COCOA's scatterer. \textcolor{black}{The distributions have been obtained by generating point-like interactions corresponding to 0.1~MeV (white circles) and 1~MeV (black circles) energy depositions uniformly inside the scatterer volume.} Dashed lines correspond to a double Gaussian fits.}\label{fig:spatial_scatterer}
\end{figure*}

\begin{figure*}[ht!]
\centering
\includegraphics[width=0.98\linewidth]{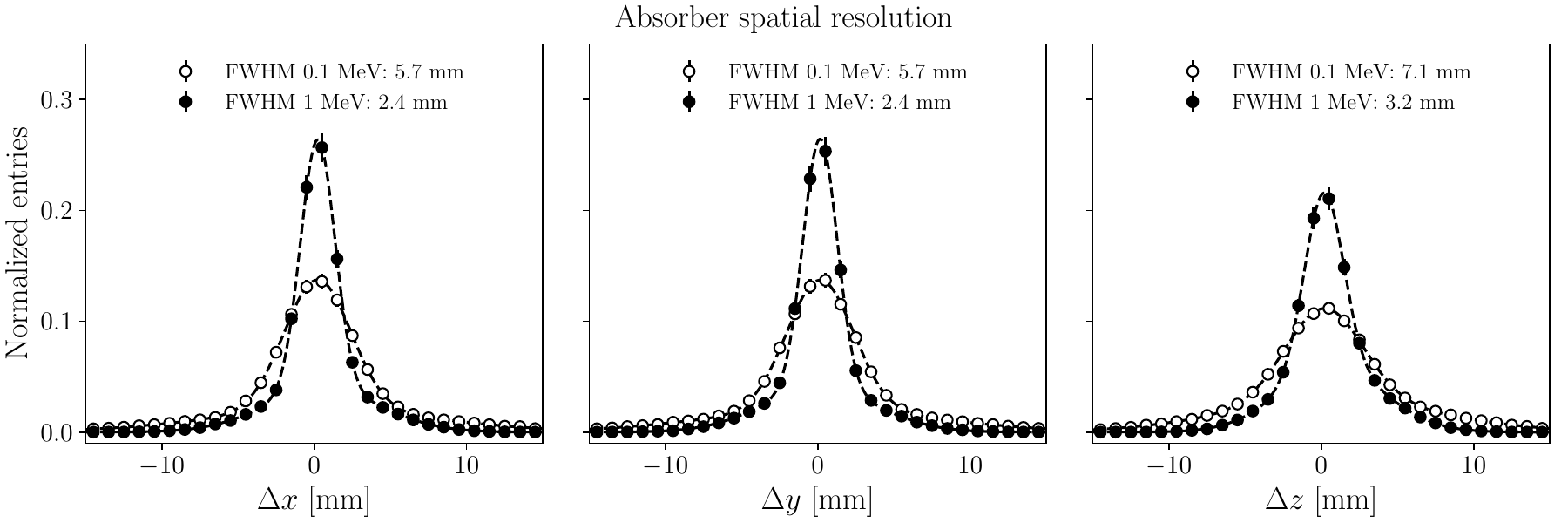}
\caption{Spatial resolution in the three dimensions for COCOA's absorber. \textcolor{black}{The distributions have been obtained by generating point-like interactions corresponding to 0.1~MeV (white circles) and 1~MeV (black circles) energy depositions uniformly inside the absorber volume.} Dashed lines correspond to a double Gaussian fit.}\label{fig:cnn_results}
\end{figure*}

\section{Performances and sensitivity}\label{sec:performances}
In order to evaluate the performances and the sensitivity of the COCOA experiment, a simulation of the detector was implemented with the Geant4 simulation toolkit~\cite{GEANT4:2002zbu}. 

\textcolor{black}{At this conceptual stage, the simulation includes only the active detector components and does not incorporate passive materials such as mechanical support structures, SiPM housings, readout electronics, thermal shielding, or micrometeorite protection. This choice was made because an accurate mass model and a detailed engineering drawing are not available at this stage of development, and the goal of the current study is to assess the intrinsic performance of the proposed configuration. We acknowledge that these elements will attenuate incoming gamma rays and affect the background rate and sensitivity estimates. These effects will be addressed in future work, once a more complete instrument model becomes available.} 

For the treatment of the Compton scattering, the Monash University model~\cite{Brown:2014gza} was adopted, which is dedicated to low-energy (below 20~MeV) simulations and is implemented in the \texttt{G4LowEPComptonModel} class. Optical photons were generated both in the scatterer and in the absorber using the \texttt{UNIFIED}~\cite{levin1996more} optical model. 

The scintillator is defined as a mix of LAB/PPO at 80\% wt and paraffin wax at 20\% wt. Its light yield is set at 8,000 photons/MeV, 20\% lower than the nominal LAB/PPO of 10,000 photons/MeV due to the presence of wax. However, the wax percentage can be in principle reduced while maintaining opacity, increasing the light yield.  The absorption length was assumed to be the same as that of a traditional LAB/PPO cocktail, using values from ref.~\cite{zhang2020complete}. The Rayleigh scattering length of the NoWaSH is set at 5~mm, although it could in principle be reduced up to 0.5~mm~\cite{Bezerra:2024frp}, improving the spatial resolution. The walls of the scatterer volume are assumed to be perfect light absorbers. 

The material of the absorber is CsI(Tl) with a light yield of 50,000 photons/MeV. The reflectivity of the ESR Vikuiti film used as wrapping is set at 98\% and is assumed to be perfectly specular (no Lambertian component).

\subsection{Spatial resolution}\label{sec:spatial_res}
\subsubsection*{Scatterer}
\textcolor{black}{The Monte Carlo simulation of the COCOA scatterer shows that the LiquidO technology is able to achieve sub-cm resolution in the transverse dimensions $(x,y)$ and approximately $1.5$~cm resolution in the longitudinal dimension $z$, consistent with the results of the prototype of ref.~\cite{LiquidO:2024piw}}.  

Figure~\ref{fig:spatial_scatterer} shows the difference between the true and reconstructed interaction points for two datasets of $10^5$ scintillation photons each. The datasets correspond to energy depositions of 1~MeV and 0.1~MeV, respectively, with events generated uniformly throughout the scatterer volume. A fiducial cut of 2~cm from each side of the volume has been applied. In order to obtain the spatial coordinates of the interaction point, the charge-weighted centroid of the fibers positions is calculated for both the U and V planes. Then, the reconstructed $(x,y,z)$ coordinates are defined as the charge-weighted point of closest approach between two lines, each passing through one centroid and inclined at $\pm12^{\circ}$, respectively.

Resolution along $x$ is slightly worse than along $y$, since the fibers have alternate inclinations in that direction, smearing the response. Although the resolution along the longitudinal dimension $z$ is relatively worse, fiber inclination and pitch can be further optimized and more advanced reconstruction algorithms (e.g., neural networks) could be applied in future studies. %Also, significant improvements could be achieved by arranging the fibers in a grid for each vertical plane, as in ref.~\cite{LiquidO:2024piw}. However, this approach would increase the number of channels (at parity of pitch), and thus has been excluded from the baseline design of COCOA.

\subsubsection*{Absorber}
In a pixelated crystal calorimeter, the reconstruction of the interaction vertex usually requires a single pixel fired in the detector unit.
In the case of COCOA, this straightforward algorithm cannot be applied, since all the 64 SiPMs reading out the crystal will generate a signal when a gamma interacts. 

Thus, in order to reconstruct the position of the interaction vertex, we implemented a reconstruction algorithm based on convolutional neural networks (CNNs), which have been widely adopted for image recognition tasks with extraordinary success~\cite{li2021survey, yamashita2018convolutional}. 

\begin{figure}[ht!]
\centering
\includegraphics[width=.99\linewidth]{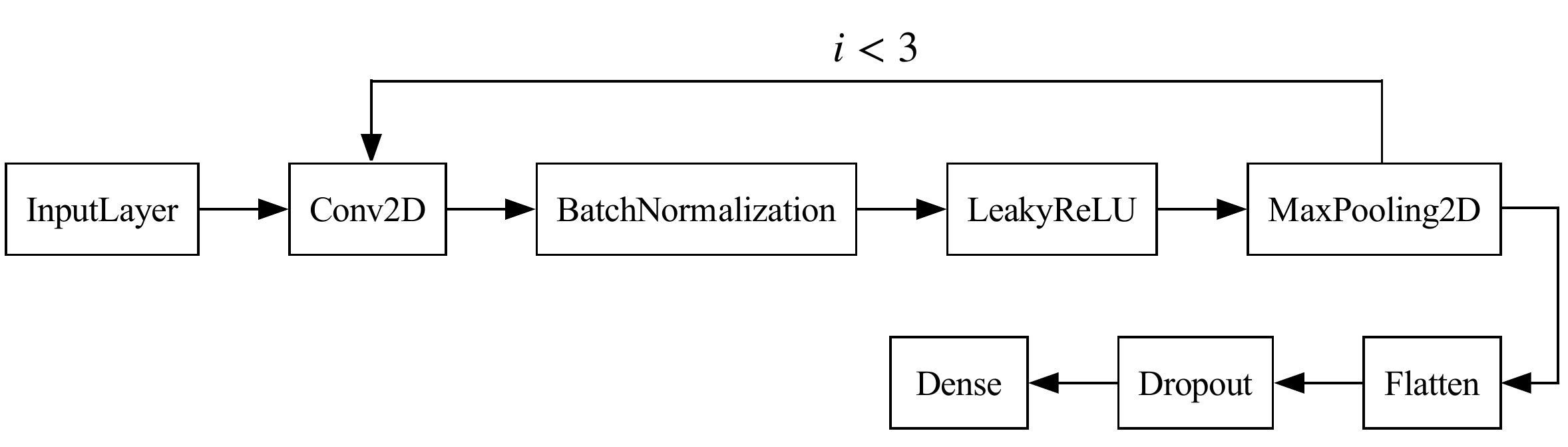}
\caption{Model of the neural network used to predict the position of the interaction vertex inside the crystal, taking as input a heatmap of the photoelectrons detected by each SiPM. The convolutional block is repeated three times and then fed to a dense layer, after passing through a flatten and a dropout layer.}\label{fig:cnn_scheme}
\end{figure}

The full network is made of three consecutive convolutional blocks (Conv2D + BatchNorm + LeakyReLU + MaxPooling2D) followed by a Flatten, a Dropout, and a Dense layer, as shown in fig.~\ref{fig:cnn_scheme}. The network was implemented in PyTorch~\cite{pytorch} and trained for 10 epochs on $10^7$ 1-MeV and 0.1-MeV events. Fig.~\ref{fig:cnn_results} shows the difference between the true interaction point and the one predicted by the network: millimeter-scale resolution is achieved in all three directions. \textcolor{black}{This result is consistent with measurements reported in the literature, mainly from experiments focused on PET scanner development~\cite{li2008high, sanaat2020depth, gonzalez2019novel}}.

\subsection{Angular resolution}\label{sec:arm}
\subsubsection*{Compton scattering}
The angular resolution $\sigma_{\theta}$ for Compton scattering events can be estimated from the spatial and energy resolutions using the following equation~\cite{ODAKA2007878}:
\begin{equation}
    \sigma^2_{\theta} = \delta\theta_E^2+\delta\theta_r^2+\delta\theta_{DB}^2.
\end{equation}
The energy resolution directly impacts the precision of the Compton scattering angle measurement, thus contributing with a factor $\delta\theta_E$. The magnitude of this effect can be calculated through error propagation from eq.~\eqref{eq:compton}. The Mini-LiquidO prototype showed that an optimized LiquidO detector would be able to achieve a light yield of up to 500 photoelectrons/MeV~\cite{LiquidO:2025qia}, which corresponds to an energy resolution of approximately $\sigma_s = 5\%/\sqrt{\mathrm{MeV}}$~\cite{LiquidO:2019mxd}. LaBr$_3$(Ce) has an energy resolution of approximately 3\% FWHM at 662~keV~\cite{shah2002labr}, while CsI(Tl) can reach values below 5\% FWHM~\cite{e-ASTROGAM:2016bph, Hawrami:2021xdm} at the same energy. Thus, we conservatively set the energy resolution of the absorber at $\sigma_a = 2.5\%/\sqrt{\mathrm{MeV}}$.

Uncertainty in the position measurements affects the axis of the Compton cone. It can by estimated as $\delta\theta_r \lesssim \tan(\Delta x/D)$, where $\Delta x$ is the spatial resolution and $D$ is the distance between interactions. 

There is also an irreducible component which fundamentally limits the resolution that can be achieved by a Compton telescope. In the scatterer, electrons are bound in an atom with a certain momentum, which adds an additional uncertainty $\delta\theta_{DB}$ to the angular resolution~\cite{zoglauer2003doppler}. This effect, called \emph{Doppler broadening} is in general smaller for low-Z materials, such as liquid scintillators. 

In this context, the performances of a Compton telescope are typically quantified by the Angular Resolution Measure (ARM). This is defined as the difference between the kinematically calculated Compton scatter angle $\theta_{\mathrm{kin}}$, obtained from eq.~\eqref{eq:compton} and the geometrically calculated Compton scatter angle $\theta_{\mathrm{geo}}$:
\begin{equation}
     \theta_{\mathrm{geo}} = \arccos\left( \frac{\vec{g_0}\cdot\vec{g_1}}{|\vec{g_0}||\vec{g_1}|} \right).
\end{equation}
Here $\vec{g_0}$ is the initial direction of the gamma, which in our case is taken from simulation, and $\vec{g_1}$ is the scattered one, which is affected by the spatial resolution of the detector.

\begin{figure}[ht!]
\centering
\includegraphics[width=.99\linewidth]{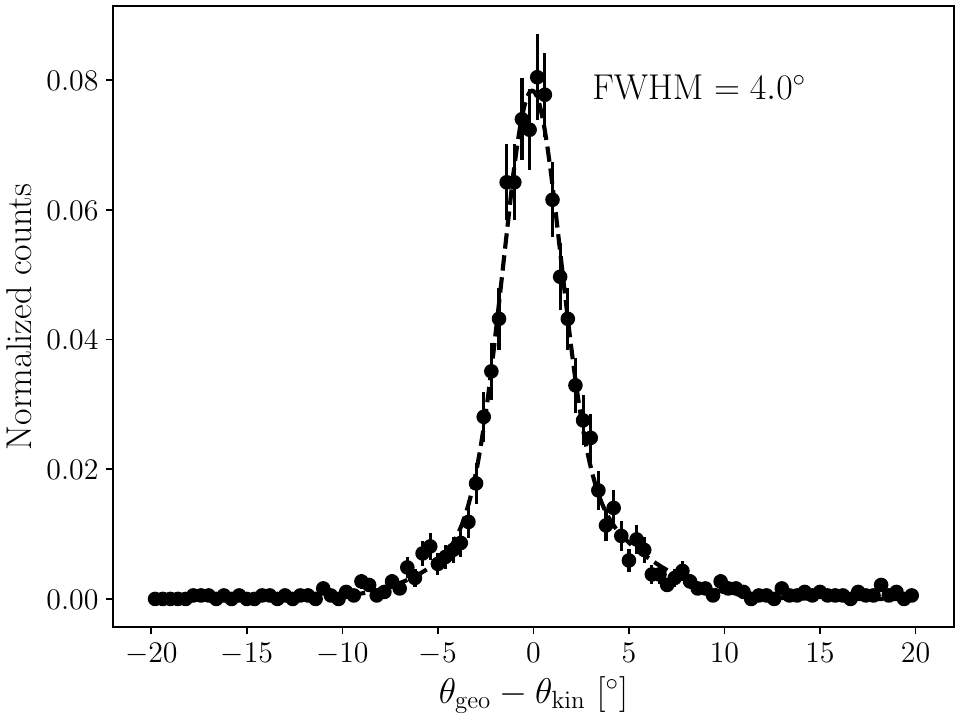}
\caption{ARM for on-axis 1~MeV gamma rays producing two-site events in the COCOA detector, obtained with a Geant4 simulation. The dashed line corresponds to a double Gaussian fit with $\mathrm{FWHM}=4.0^{\circ}$.}\label{fig:arm}
\end{figure}

A Geant4 simulation shows that the ARM for on-axis 1~MeV gamma rays producing two-site events in the COCOA detector has a resolution of approximately 4.0$^\circ$ FWHM (see fig.~\ref{fig:arm}), \textcolor{black}{which is approximately 30\% better than the one obtained by COMPTEL}~\cite{schonfelder1993instrument}. %Although challenging at low energies, the LiquidO technology may also enable the reconstruction of the direction of the recoil electron, produced in the Compton scattering interaction. In these cases, momentum conservation can further restrict the direction of the gamma ray, improving the angular resolution~\cite{akyuz2004enhanced}. The magnitude of this improvement will be quantified in future studies.

\subsubsection*{Pair production}
The LiquidO technology is able to combine topological information with the time structure of different energy depositions for the same event (its \emph{energy flow})~\cite{anatael_cabrera_2022_6697273}, enabling powerful particle identification capabilities~\cite{LiquidOConsortium:2023bqe}. \textcolor{black}{It is also worth noting that topological information alone, simple \emph{light-ball counting}, can effectively distinguish between electrons (which typically produce a single light ball) and positrons (which yield multiple smaller light balls due to annihilation gammas), as discussed in ref.~\cite{LiquidO:2019mxd}. }

In addition, hybrid solutions, such as the ones based on WbLS or slow fluors~\cite{Dunger:2022gif}, can further increase the particle discrimination power by measuring the Cherenkov/scintillation (C/S) ratio~\cite{Theia:2019non}. This is because, for positrons, the total amount of scintillation light in the detector includes the $2\cdot511$~keV energy from the annihilation gammas, which produce very little Cherenkov light~\cite{NuDoubt:2024jax}. At parity of deposited energy, then, the amount of Cherenkov light is lower with respect to an electron event. \textcolor{black}{While the detector described in ref.~\cite{LiquidO:2025qia} was not optimized to separate Cherenkov and scintillation signals, other configurations, such as those employing WbLS or slow fluors, can achieve C/S ratios of approximately 10\%~\cite{NuDoubt:2024jax}. Furthermore, a \emph{light-ball-dependent} C/S ratio enables the identification of the primary vertex in complex topologies, providing an additional handle for particle identification~\cite{NuDoubt:2024jax}.}

\begin{figure}[ht!]
\centering
\includegraphics[width=.99\linewidth]{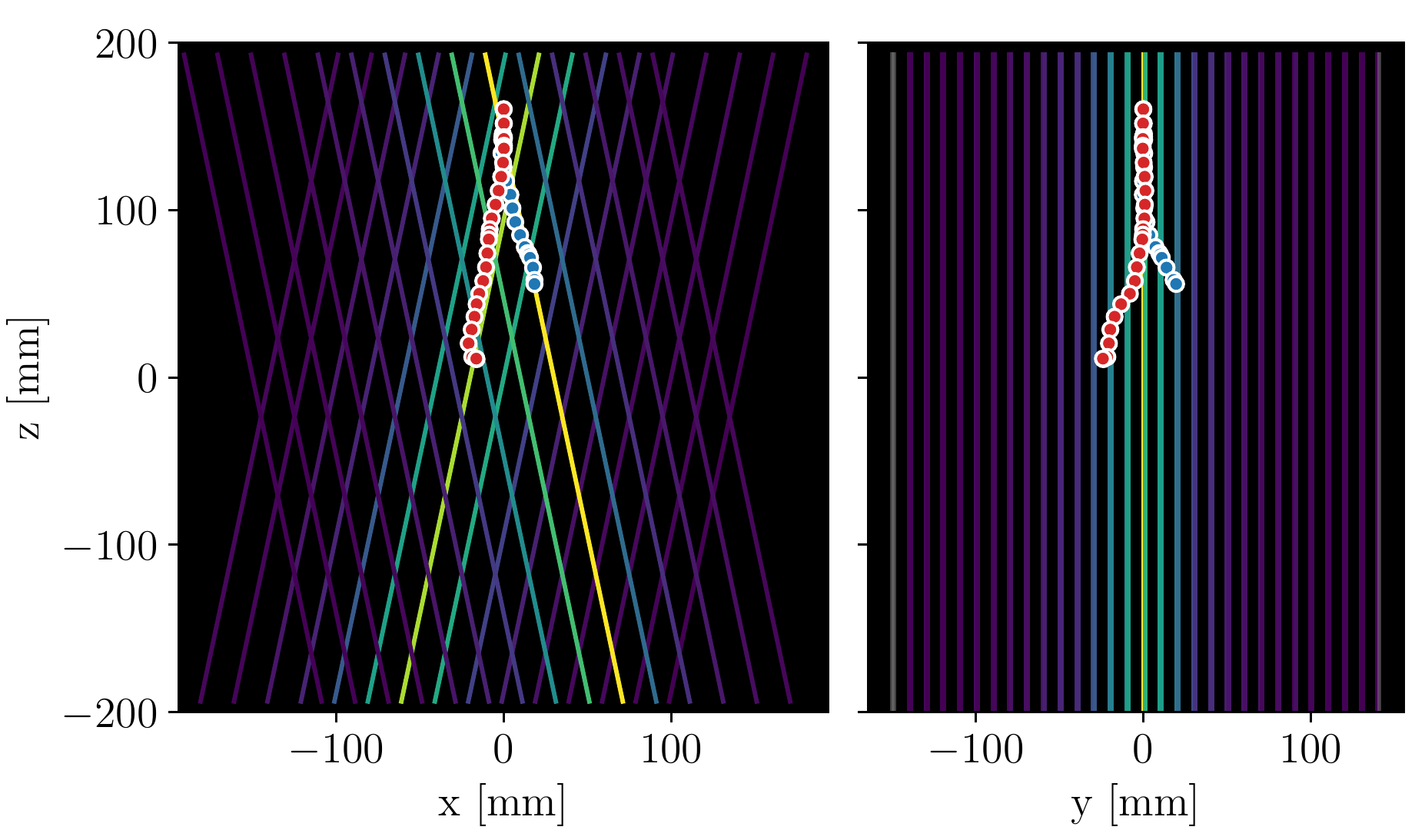}
\caption{Two-dimensional projections of a pair-production event in the COCOA scatterer volume. The red and blue dots correspond to the energy depositions of the electron and positron, respectively.}\label{fig:pair_evd}
\end{figure}

Thus, for pair-production interactions, it might be possible to distinguish between electrons and positrons, as proposed also in ref.~\cite{Grassi:2018hew}, and directly measure their directions from the trail of energy depositions in the detector, as shown in fig.~\ref{fig:pair_evd}. The performance of the reconstruction algorithm will greatly depend on several parameters, such as the relative orientation between gamma ray and fiber planes, the collinearity of the $e^+e^-$ pair, and the time resolution of the detector. As a first approximation, the positions of the energy depositions have been smeared according to the spatial resolution obtained in section~\ref{sec:spatial_res} and a principal component analysis is performed on the first 5~cm of the two trails. Thus, the direction of the gamma ray is estimated as the energy-weighted average of the reconstruction directions of the electron and of the positron. 

For this category of events, the angular resolution is defined as the angular difference between true direction and reconstructed one that contains 68\% of the events.

\begin{figure}[ht!]
\centering
\includegraphics[width=.99\linewidth]{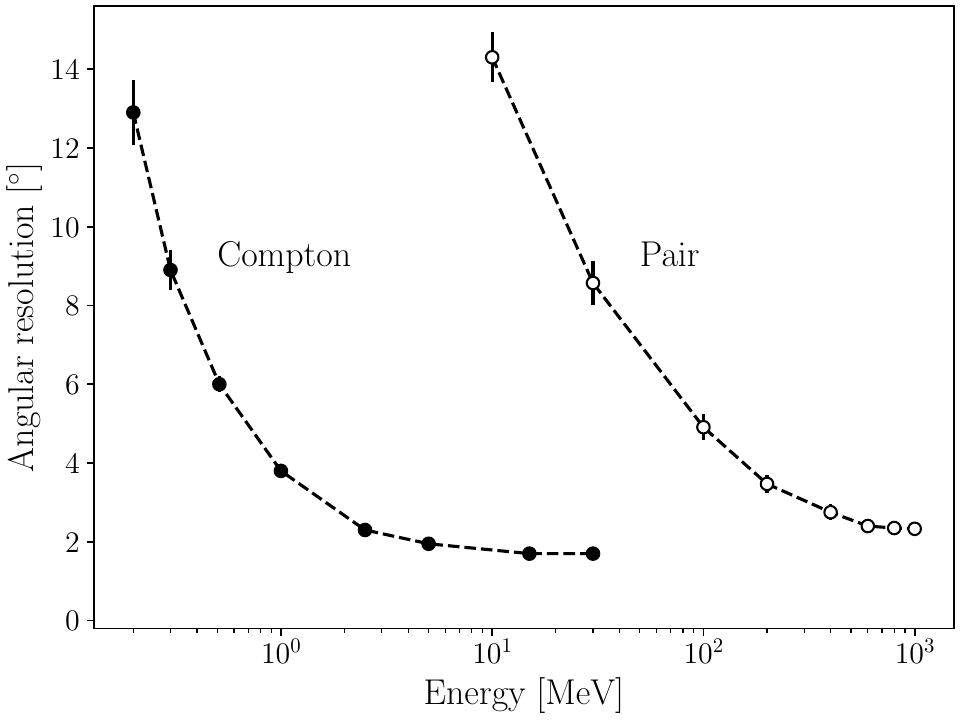}
\caption{Angular resolution for on-axis gammas as a function of the energy in the COCOA detector for Compton (\textcolor{black}{FWHM of the ARM}, black dots) and pair-production (\textcolor{black}{angular difference that contains 68\% of the events}, white dots) events.}\label{fig:arm_energy}
\end{figure}

Figure~\ref{fig:arm_energy} shows the angular resolution as a function of the energy for on-axis gammas, for both Compton and pair-production interactions. Its value approaches approximately 2$^\circ$ in both categories at high energies. 

\subsection{Effective area}
The effective area of a gamma-ray telescope is the equivalent area of a perfect detector that would intercept the same number of gamma rays as the actual telescope~\cite{Fleischhack:2021mhc}. It is defined as:
\begin{equation}
    A_{\mathrm{eff}} = A_{\mathrm{sim}}\cdot\frac{N_{\mathrm{reco}}}{N_{\mathrm{sim}}},
\end{equation}
where $A_{\mathrm{sim}}$ is the area of the surface from which the simulated gammas are thrown, $N_{\mathrm{reco}}$ is the number of reconstructed events, and $N_{\mathrm{sim}}$ is the number of simulated events. The result does not depend on $A_{\mathrm{sim}}$, as long as its value is large enough to cover the field of view of the detector.

It has been estimated both for Compton scattering and pair-production events with a dedicated Geant4~\cite{GEANT4:2002zbu} simulation. 

\textcolor{black}{In the scatterer, the simulation output corresponds to the charge collected by the SiPMs coupled to the WLS fibers. For Compton events, these charge distributions are clustered using a charge-weighted mean shift algorithm~\cite{fukunaga1975estimation} (see fig.~\ref{fig:event_scatterer} for an example). Events are discarded if the number of reconstructed clusters in the U and V planes does not match. Each cluster in one plane is associated with the cluster in the other plane that has the closest total charge. The centroids of the matched clusters in the U and V planes are then used to reconstruct the interaction vertex, following the same procedure described in section~\ref{sec:spatial_res}. This reconstruction approach is preliminary and intended to provide a first-order evaluation of the detector's spatial performance.}

\textcolor{black}{In the absorber, the energy deposition is assumed to result from photoabsorption and is therefore treated as point-like. The heatmap corresponding to the charge collected by the SiPMs is fed to the neural network described in section~\ref{sec:spatial_res}, which returns a three-dimensional position.}

Then, the following selection criteria are applied: (1) there must be energy deposited in the scatterer (2) the total reconstructed energy deposited in the detector must fall within a $2\sigma$ window around the photopeak. In the case of Compton scattering, (3) there must be energy deposited in the absorber, (4) the number of Compton interactions, estimated from the number of reconstructed clusters in the scatterer, is limited to a maximum of three, and (4) the distance between scattering points must be larger than 5~cm. \textcolor{black}{This procedure provides an approximate estimate of the effective area; a more precise determination would require more advanced reconstruction algorithms, which are beyond the scope of this paper.}

\begin{figure}[htbp]
\centering
\includegraphics[width=.99\linewidth]{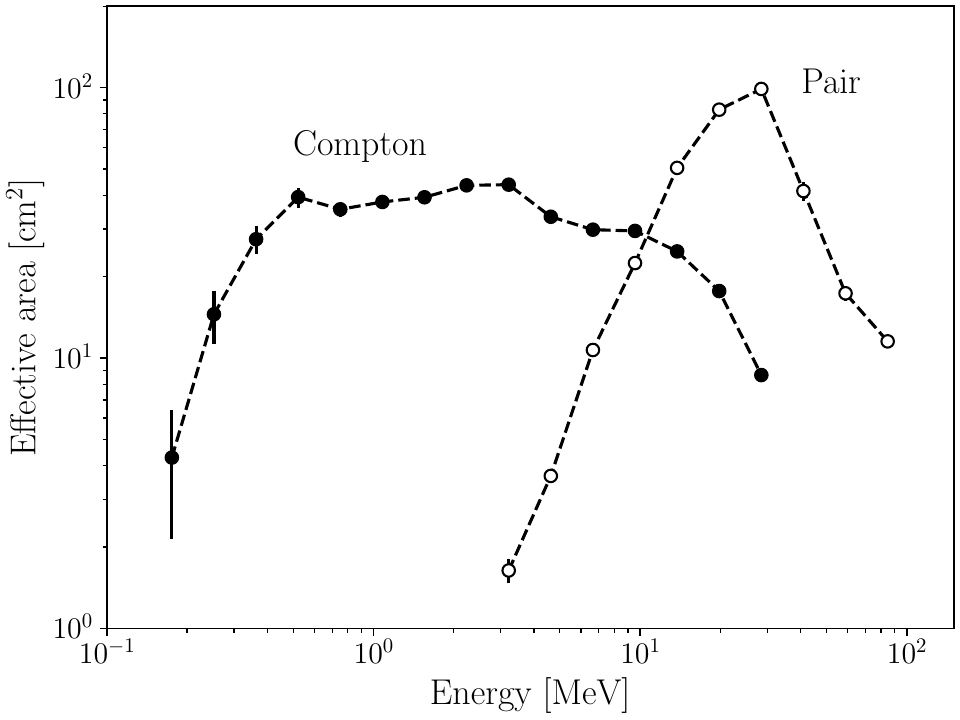}
\caption{Effective area of the COCOA telescope for Compton (black dots) and pair-production (white dots) events as a function of the gamma ray energy. \textcolor{black}{The value used for the sensitivity calculation corresponds to the sum of the two components.}}\label{fig:effective}
\end{figure}

Figure~\ref{fig:effective} shows the effective area for COCOA as a function of the gamma energy, \textcolor{black}{obtained by simulating $10^5$ gamma rays uniformly distributed between 0.2 and 100~MeV}. As expected, for Compton interactions it is mostly constant above 0.5~MeV and starts decreasing at $\mathcal{O}(10~\mathrm{MeV})$, where pair production begins to be dominant. \textcolor{black}{Compared with COMPTEL~\cite{schonfelder1993instrument}, the total effective area is approximately 4 times higher at 1~MeV and 1.5 times higher at 10~MeV.}

Notably, the capability of the LiquidO technology to reconstruct the position of the energy deposition within the opaque scintillator volume could allow also the reconstruction of events with 3+ interactions, using Compton kinematic discrimination~\cite{APRILE1993216, boggs2000event}. This method compares the scattering angle calculated with the Compton equation \eqref{eq:compton} with the angle calculated geometrically, considering the interaction points. In this way, photons that do not deposit all of their energy in the detector can be effectively rejected. 

For pair-production events, the effective area reaches a peak at approximately 30~MeV and then decreases, since the fraction of uncontained gammas starts becoming significant. Thus, the performances of COCOA at $\mathcal{O}(100~\mathrm{MeV})$ could be improved by adding extra layers of scintillating crystals, increasing the total number of radiation lengths in the absorber.

\subsection{Time-of-flight}\label{sec:tof}
\begin{figure}[htbp]
\centering
\includegraphics[width=.99\linewidth]{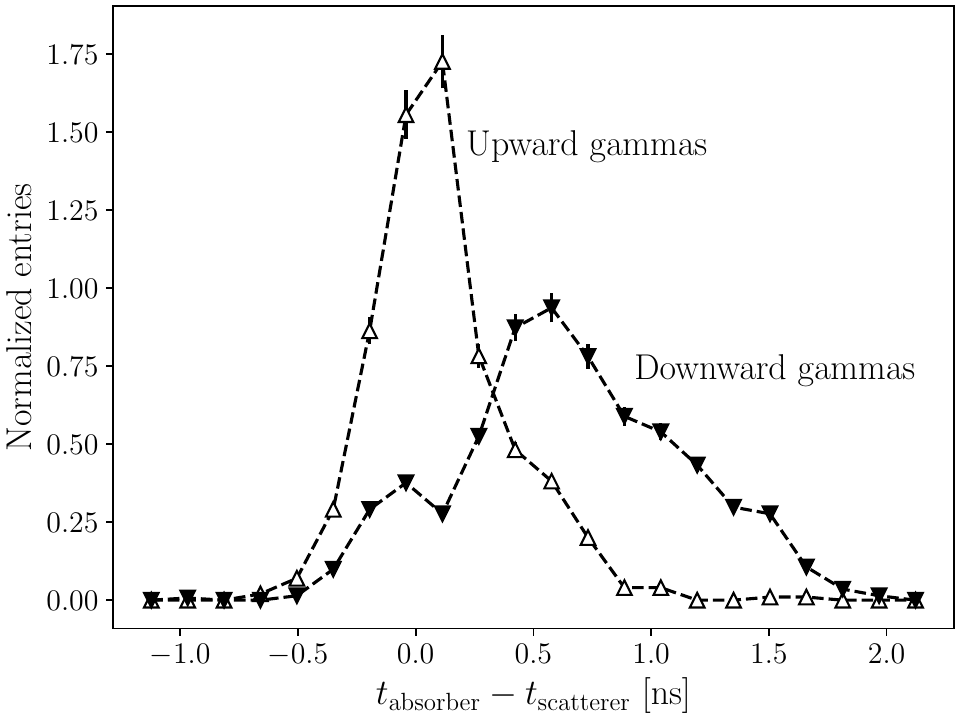}
\caption{\textcolor{black}{Time difference between the first photoelectron in the absorber and the first photoelectron in the scatterer for downward (black triangles) and upward  (white triangles) gammas with an energy of 5~MeV.}}\label{fig:tof}
\end{figure}

The COMPTEL detector was able to measure the TOF between the interaction in the scatterer and the one in the absorber, which provided a strong discriminator against instrumental background~\cite{comptel_tof}. In more compact detectors like COCOA, TOF discrimination is more challenging due to the reduced distance between interaction points.  Thus, in order to assess the TOF capabilities of COCOA, we simulated upward- and downward-going gammas in two transverse planes, the former below the absorber and the latter above the scatterer, impinging the detector perpendicularly. The simulation took into account the light response of the scatterer, as measured in ref.~\cite{LiquidO:2025qia}, and the light response of LaBr$_3$(Ce)~\cite{Giaz:2013hna}, which is faster than the NaI(Tl) crystals used by COMTPEL. The single-p.e. resolution for the SiPMs was set to 150~ps, in agreement with existing measurements in the literature~\cite{Gundacker:2013ywa, Nemallapudi:2016qxh}. Figure~\ref{fig:tof} shows the time difference between the first photoelectron detected in the absorber and the first photoelectron detected in the scatterer for single-scatter events produced by 5 MeV gamma rays. Applying a simple threshold cut of $\Delta t > 0.25$~ns results in a 75\% rejection rate for upward-going gammas and an 80\% efficiency for downward-going gammas. Although challenging, a possible improvement in this context might be achieved by increasing the detection efficiency of Cherenkov photons. This feature require careful experimental validation and will be the focus of future work.

\subsection{Field of view}\label{sec:fov}
\textcolor{black}{The detector's 3D capabilities in principle allow for a very wide field of view (FOV), as gamma rays are not limited to entering only through the top face. However, the detection efficiency for gamma rays close to the horizon is low, since the gamma traverses a smaller amount of material and, in the case of Compton events, must have a large scattering angle in order to hit the absorber. Thus, the COCOA FOV is evaluated from the angular dependence of the sensitivity, following the prescription of ref.~\cite{e-ASTROGAM:2016bph}. The average half width at half maximum (HWHM) of the inverse of the sensitivity distribution as a function of the polar angle is 41$^{\circ}$. The sky coverage fraction is 32\%, corresponding to an effective FOV of approximately 4~sr.}

\subsection{Sensitivity}
The sensitivity of a Compton telescope is its most crucial performance characteristic. It represents the minimum detectable source flux and its determination requires a careful estimation of the backgrounds and the detector's performances.

The following approximation can be used to estimate the gamma-ray continuum sensitivity $S_k(E)$ for a $\sigma$ confidence level $k$, assuming a background count limited observation~\cite{Aramaki:2019bpi}:

\begin{equation}
    S_k(E) \approx k\sqrt{\frac{\Phi_B\Delta\Omega}{A_{\mathrm{eff}}T_{\mathrm{eff}}\Delta E}},
\end{equation}
where $\Phi_B$ is the background flux, $A_{\mathrm{eff}}$ is the effective area, $T_{\mathrm{eff}}$ is the effective observation time and $\Delta E = 0.5E$ is the energy bandwidth around energy $E$. 

The angular resolution element on the sky map $\Delta\Omega$ is defined for Compton events as~\cite{Lucchetta:2017epw}:
\begin{equation}
\Delta\Omega(E)=2\pi\left[\left(\cos\bar{\theta}(E)-1.4\cdot\sigma_{\mathrm{ang}}\right)-\left(\cos\bar{\theta}(E)+1.4\cdot\sigma_{\mathrm{ang}}\right)\right],
\end{equation}
\textcolor{black}{where $\bar{\theta}(E)$ is the average Compton scattering angle at energy $E$}, $\sigma_{\mathrm{ang}}$ is the standard deviation corresponding to the angular resolution, and the $1.4$ factor gives an approximate optimal selection window for a background-limited measurement and a Gaussian-distributed signal. In the case of pair-production events, $\Delta\Omega$ is more straightforwardly given by the point-spread function~\cite{ackermann2013determination}.

Regarding $\Phi_B$, we adopt two different estimations, depending on whether the experiment is conducted on a balloon or a satellite, as detailed in section~\ref{sec:mission}. 

For the balloon case, the primary background is typically represented by the atmospheric photon flux, which has been estimated at the operational altitude (35~km) and latitude (65$^{\circ}$~N) using EXPACS~\cite{sato2018expacs}. \textcolor{black}{In addition, secondary cosmic-ray particles -- such as charged leptons, hadrons, and alpha particles -- can also interact with the detector, producing background events. At lower energies, a significant contribution also comes from extragalactic photons and the galactic diffuse background, which are not accounted for by EXPACS. Accurately evaluating these components would require a detailed mass model and reconstruction algorithms, which are not yet available at this stage of development. To provide an approximate estimate, these components have been inferred as the average difference between the total background and the atmospheric photon component, as evaluated for the COSI balloon flight in ref.~\cite{Gallego:2025fgj}. Given the approximations involved and the absence of atmospheric attenuation in the simulation, a systematic uncertainty of 50\% is assigned to this estimate.}

For the satellite case, two prompt components of the gamma-ray background are considered: \emph{albedo} photons, originating from Earth's atmosphere, and \emph{extragalactic} photons, as estimated in ref.~\cite{cumani2019background}. \textcolor{black}{A fraction of the albedo photons will be absorbed by the crystal calorimeter, which can act as an active veto. A thick veto shield (not currently included in the simulation) might be placed below the absorber to further suppress the albedo component, as in COSI~\cite{tomsick2019compton}. Albedo gammas could be partially rejected also with TOF measurement (see section \ref{sec:tof})}. However, as a conservative approach, the full albedo flux is considered as background.

\textcolor{black}{In addition, when operating in LEO, background contributions from cosmic-ray activation, particularly in regions containing high-$Z$ materials such as the supporting structure and absorber, can be the dominant component. As for the balloon case, a precise quantification would require a detailed mass model of the detector. For the purpose of obtaining an order-of-magnitude estimate, we adopt the cosmic-ray activation background evaluated for e-ASTROGAM in ref.~\cite{cumani2019background}, under the assumption of a 0$^{\circ}$ orbit inclination and neglecting the contribution from the South Atlantic Anomaly. For a smaller satellite like COCOA, the instrumental background is expected to be lower, so a scaling by mass$^{1/3}$~\cite{lucchetta2022introducing} has been applied. Given the approximations that went into this calculation, a systematic uncertainty of 50\% is assigned to the background estimate.}

\begin{figure}[htbp]
\centering
\includegraphics[width=0.99\linewidth]{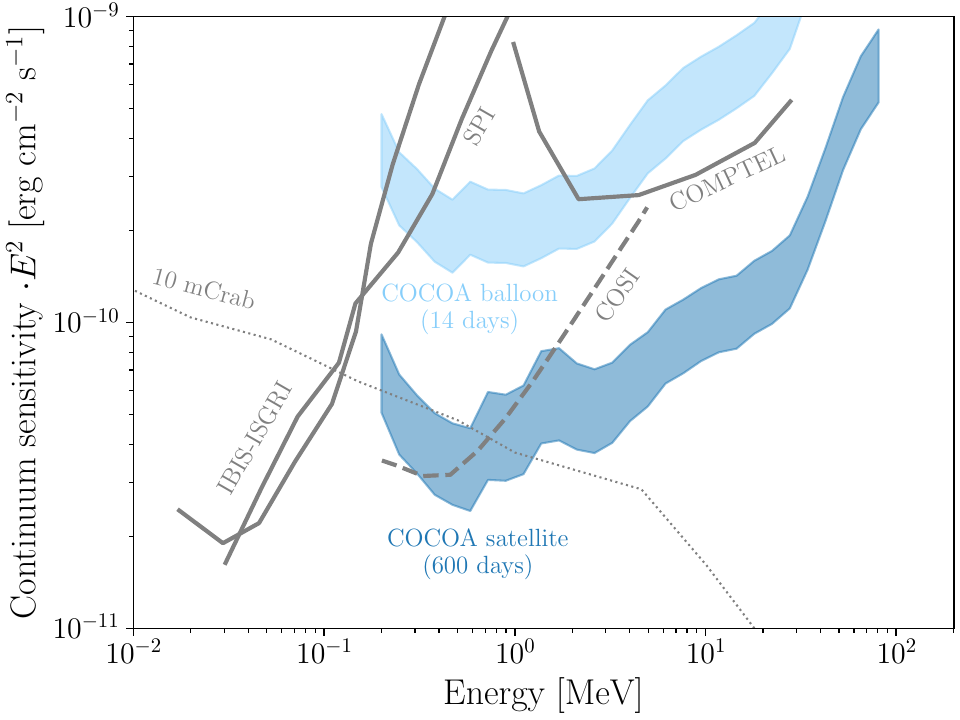}
\caption[a]{Continuum sensitivities at $3\sigma$ level of the COCOA experiment for a balloon (light blue) and a satellite (dark blue) campaign, compared with existing limits from the SPI ($10^6$~s)~\cite{roques2003spi}, COMPTEL (9 years)~\cite{schonfelder2004lessons}, and IBIS-ISGRI ($10^5$~s)~\cite{ubertini2003ibis} experiments (in gray). \textcolor{black}{The filled bands represent the $1\sigma$ systematic uncertainty}. Sensitivity requirement for COSI~\cite{tomsick2019compton} corresponds to the dashed gray line. The dotted line corresponds to the sensitivity for a gamma flux level of 10 mCrab~\cite{takahashi2013multiwavelength}.}\label{fig:mev_gap}
\end{figure}

Both atmospheric and cosmic-ray activation neutrons can also contribute to the background in gamma experiments. A naive calculation based on the neutron $\mathrm{TOF}$ between the scatterer and absorber demonstrates that a relaxed time cut of $\mathrm{TOF} < 10$~ns can reject approximately 95\% of slow neutrons (with energies below 1~MeV). Additionally, LAB-based scintillator mixtures show good fast neutron/gamma separation through pulse-shape discrimination~\cite{li2011timing}. However, the performance of such discrimination in a LiquidO detector has not yet been evaluated and a detailed quantification of the neutron background-suppressing capabilities of COCOA will be addressed in future studies.

The continuum sensitivity at $3\sigma$ level for COCOA is shown in fig.~\ref{fig:mev_gap} for two scenarios, a two-week run ($T_{\mathrm{eff}}=14$~days) with a balloon and a two-year run ($T_{\mathrm{eff}}=600$~days) with a satellite, \textcolor{black}{assuming in this case a daily all-sky scan and an instantaneous sky coverage of 32\% (see section~\ref{sec:fov})}. Even with its limited dimensions, COCOA might have the potential to improve current limits in the MeV energy band.

\begin{figure}[htbp]
\centering
\includegraphics[width=0.95\linewidth]{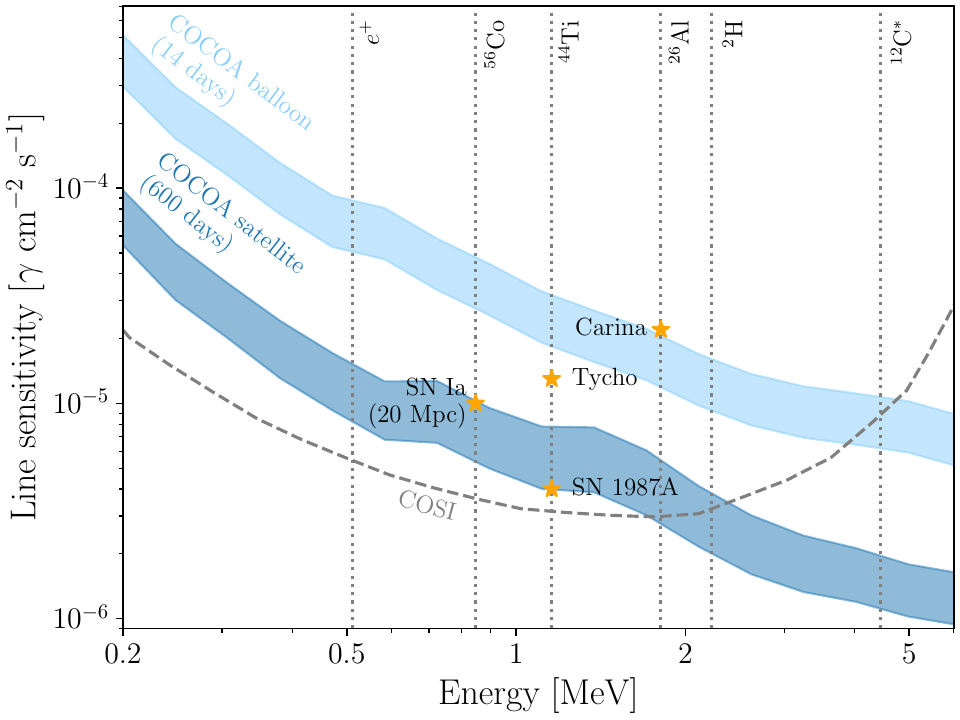}
\caption{Gamma line sensitivity of the COCOA experiment for a balloon (light blue) and a satellite (dark blue) campaign, compared with the required sensitivity from COSI~\cite{tomsick2019compton} (gray dashed line). \textcolor{black}{The filled bands represent the $1\sigma$ systematic uncertainty}. The vertical dotted lines correspond to energies of isotopes of astrophysical interest. Various notable astronomical sources are shown as yellow stars.}\label{fig:line_sensitivity}
\end{figure}

In the case of specific gamma lines, the sensitivity can be approximated as:

\begin{equation}
    S_k(E) \approx k\sqrt{\frac{\Phi_B\Delta\Omega\Delta E}{A_{\mathrm{eff}}T_{\mathrm{eff}}}},
\end{equation}
where in this case the energy bandwidth is set as $\Delta E=3\sigma_E$, with $\sigma_E$ being the standard deviation corresponding to the energy resolution at energy $E$. Fig.~\ref{fig:line_sensitivity} shows the sensitivity of COCOA for several sources of astrophysical interest: positron annihilation (0.511~MeV), Co-56 (0.847~MeV) from type Ia supernovae, Ti-44 (1.157~MeV) and Al-26 (1.809~MeV) from core-collapse supernovae, H-2 from neutron capture (2.223~MeV) and C-12$^*$ (4.438~MeV) from cosmic-ray interactions. %COCOA is competitive with the sensitivity requirement of COSI~\cite{tomsick2019compton} and is one order of magnitude better at $E>5$~MeV if deployed as satellite.

\section{Mission profile}\label{sec:mission}
The payload weight and the experiment size are compatible with Ultra Long Duration Balloon (ULDB) flights~\cite{smith2002nasa}, which can reach an altitude of approximately 35~km for a duration of 4-6 weeks. This approach is considered relatively low risk, as several Compton telescopes have previously been successfully deployed on scientific balloon missions~\cite{aprile2008compton, kierans20172016}.

\begin{figure}[ht!]
\centering
\includegraphics[width=0.95\linewidth]{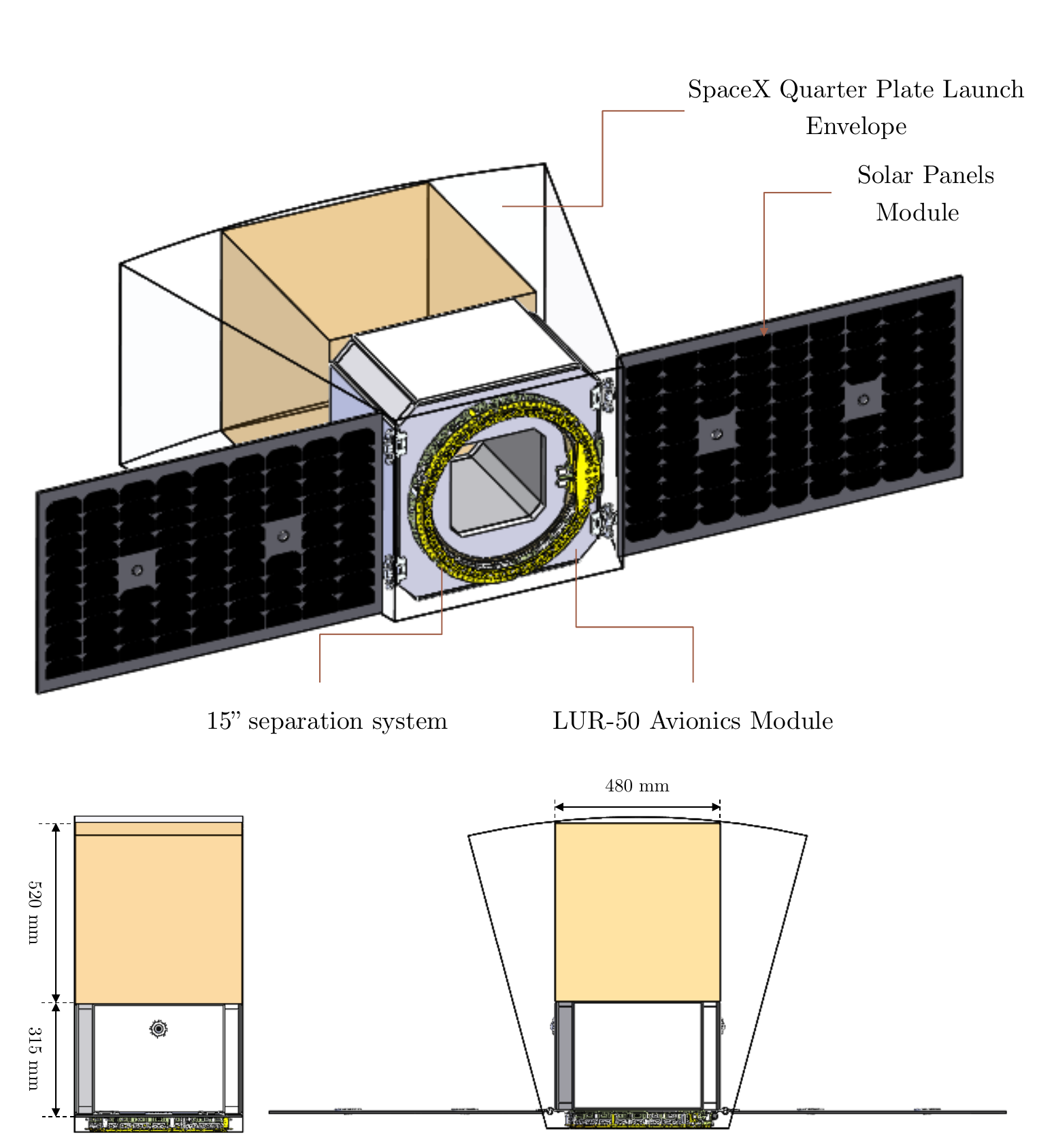}
\caption{Satellite layout in the LUR-50 configuration proposed for the COCOA detector. The module is able to provide up to 210~W of power and has 120~L of available payload volume. It includes the StarTracker ADCS system, which provides 30" attitude determination accuracy. Courtesy of AVS.}\label{fig:lur50}
\end{figure}

However, the recent dramatic decrease in the \textcolor{black}{launch} costs of LEO payload~\cite{jones2018recent} makes COCOA an attractive candidate for a satellite mission. In particular, the LUR SmallSat platform~\cite{lur}, in the proposed LUR-50 configuration, can provide 120~L of available payload volume, 210~W peak power and 20~GB download per day, largely satisfying COCOA's requirements (see fig.~\ref{fig:lur50}). The satellite could be loaded in the fairing of a SpaceX Falcon 9 launcher and placed in an equatorial LEO orbit (approximately 550 km altitude and inclination $<5^{\circ}$). %In this way, the experiment would be able provide sensitivities in the MeV range that are close to the ones of next-generation mission (GRAMS, COSI), at a fraction of the cost. 

\section{Conclusions}\label{sec:conclusions} 
The COCOA experiment has the potential to significantly advance astrophysical gamma-ray observation, focusing on the 0.2--100~MeV range and addressing the longstanding ``MeV gap" in this domain. Its design, featuring a LiquidO scatterer and a crystal calorimeter absorber, achieves spatial, energy, and angular resolutions that are comparable with most modern Compton telescopes, all within a compact framework.

COCOA's scalability and reduced channel count make it a versatile candidate for deployment as either a balloon payload or a microsatellite.
\textcolor{black}{Simulations studies suggest that COCOA could improve COMPTEL and SPI gamma continuum sensitivity, depending on the energy and the mission type (balloon or satellite). More detailed detector simulations and background estimations are required to precisely quantify the magnitude of the improvement. Increasing the sensitivity by 1-2 orders of magnitude is crucial, since it would allow to clarify the nature of the COMPTEL excess~\cite{tsuji2023mev}}. In addition, although this proposal focuses on detection sensitivity, COCOA might be able to also characterize the gamma polarization, by measuring the azimuthal scattering angle.

\section*{Acknowledgments}
The authors are thankful to M. Angel Carrera and R. Diaz de Cerio (AVS) for providing us information about the LUR platform, and to S. Bonoli, F. Monrabal, A. Simón Estévez (DIPC), A. Zoglauer (UC Berkeley) for the invaluable feedback.

We acknowledge the support from the CNPq/CAPES in Brazil, the McDonald Institute in Canada, the Charles University in the Czech Republic, the CNRS/IN2P3 in France, the INFN in Italy, the Fundação para a Ciência e a Tecnologia (FCT) in Portugal, the CIEMAT, the ``la Caixa Foundation'' (ID 100010434, code LCF/BQ/PI22/11910019) in Spain, the STFC/UKRI/Royal Society in the UK, the University of California at Irvine, Department of Defense, Defense Threat Reduction Agency (HDTRA1-20-2-0002) and the Department of Energy, National Nuclear Security Administration, Consortium for Monitoring, Technology, and Verification (DE-NA0003920), Brookhaven National Laboratory supported by the U.S. Department of Energy under contract DE-AC02-98CH10886 in the USA for their provision of personnel and resources.


\bibliographystyle{model1-num-names}
\bibliography{biblio}

\end{document}